\documentclass[12pt,aos]{article}
\usepackage{mathrsfs}
\usepackage{epsf}
\usepackage{amsthm,amsmath,amssymb,graphicx,graphics,epsfig}
\usepackage{lineno}
\usepackage{hyperref}
\usepackage[authoryear,round]{natbib}
\usepackage{longtable}
\usepackage{float}
\usepackage{multirow}
\usepackage{graphicx,graphics,epsfig}
\usepackage{cite}
\usepackage[graphicx]{realboxes}
\usepackage{rotating}
\def\Cov{\mbox{Cov}}

\def\diag{\mbox{diag}}

\def\bSig\mathbf{\Sigma}

\newcommand{\Rmnum}[1]{\uppercase\expandafter{\romannumeral #1}}

\begin{document}
\baselineskip 8mm \setcounter{page}{0} \thispagestyle{empty}
\begin{center}
{\Large \bf Robust approach for variable selection with high
	dimensional longitudinal data analysis} \vspace{3mm}

Liya Fu$^{1^*}$, Jiaqi Li$^{1^{**}}$ and You-Gan Wang$^{2}$\\
{\it $^1$School of Mathematics and Statistics, Xi'an Jiaotong
University, China \\
$^2$ School of Mathematics and Physics, Queensland University of
Technology, St Lucia, QLD 4072, Australia}  \\
$^*$ {\it Email:fuliya@mail.xjtu.edu.cn}  \\
$^{**}${\it Email:lijq0305@stu.xjtu.edu.cn}

\end{center}

\noindent{\bf Summary.} 
This paper proposes a new robust smooth-threshold estimating equation to select important variables and automatically estimate parameters for high dimensional longitudinal data.
A novel working correlation matrix is proposed to capture correlations within the same subject.
The proposed procedure works well when the number of covariates $p$ increases as the number of subjects $n$ increases. %and even when $p$ exceeds $n$.
The proposed estimates are competitive with the estimates obtained with the true correlation structure,
especially when the data are contaminated. Moreover, the proposed method is robust against outliers in the response variables
and/or covariates. Furthermore, the oracle properties for robust smooth-threshold estimating equations under ``large $n$, diverging $p$'' are  established under some regularity conditions.
Extensive simulation studies and a yeast cell cycle data are used to evaluate the performance of the proposed method, and
results show that our proposed method is competitive with existing robust variable selection procedures.


\newpage

\label{firstpage}
\section{Introduction}
\label{s:intro}
Longitudinal data is usually collected by repeatedly observing the results for each subject at several points in time. It has been widely used in
medical and economic research over the past decade. High-dimensional longitudinal data consisting of repeated measurements with a large number of covariates has
become increasingly common in practical application. The number of covariates can be quite large, especially when the interactions of various factors are considered.
Nevertheless, there is only a subset of covariates related to the response variables, and the redundant variables can affect the accuracy and efficiency of estimation.
Therefore, it is important to develop a new methodology to select the important variables in high-dimensional longitudinal data.

\begin{figure}
	\begin{center}
		\includegraphics[width=0.5\textwidth]{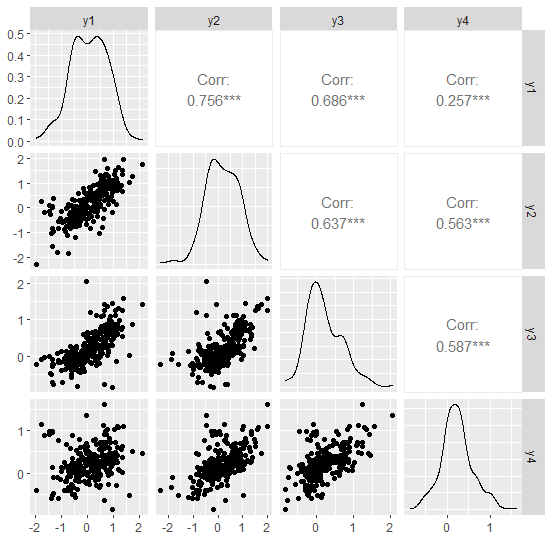}
	\end{center}
	\caption{The correlation plots of the log-transformed gene expression level. Here ``y1'' represents the first observation, and so forth.
		The correlation coefficients among factors, the density maps of them, and the scatter plots of two factors
		lie on the upper right triangle, the diagonal, and the lower left triangle, respectively.\label{ycor} }
\end{figure}

\begin{figure}
	\begin{center}
		\includegraphics[width=0.5\textwidth]{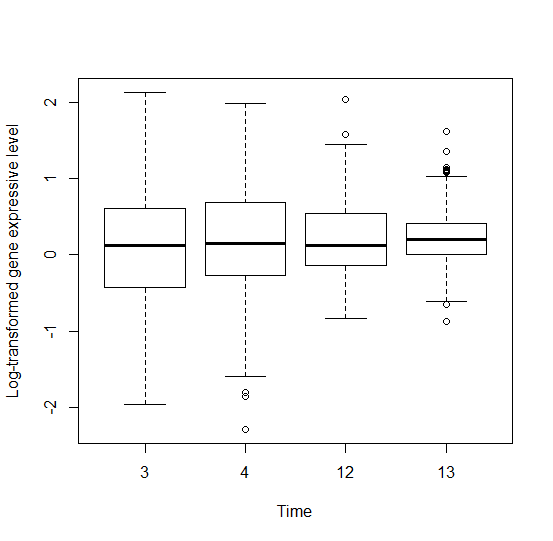}
	\end{center}
	\caption{The boxplots of log-transformed gene expression level over four time points.\label{youtlier}}
\end{figure}

\begin{figure}
	\begin{center}
		\includegraphics[width=0.5\textwidth]{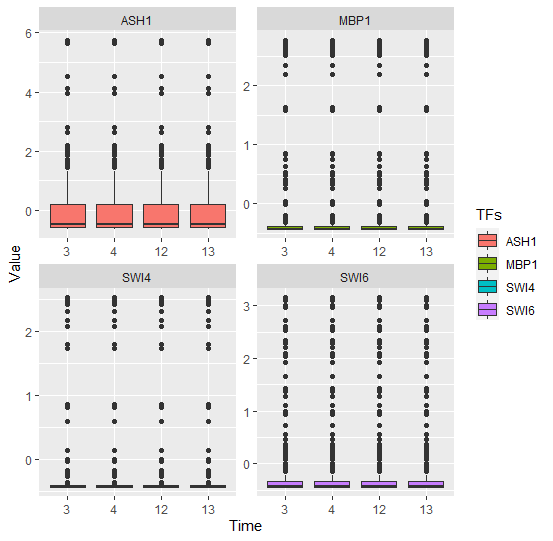}
	\end{center}
	\caption{The boxplots of four important TFs: ASH1, MBP1, SWI4, and SWI6 over four time points.\label{Xoutlier}}
\end{figure}

To select the important variables in longitudinal data analysis, \citet{pan01} proposed a quasi-likelihood
information criterion (QIC) based on an independence assumption, which can be used to select variables and working correlation matrices.
\citet{wang09} combined the Bayesian information criterion with quadratic inference function, which does not require the full likelihood or quasi-likelihood.
	However, these two methods can be
	computationally intensive when the dimension of covariates $p$ is  large.
\citet{tian14} extended the SCAD-penalized quadratic inference function to analyze semiparametric varying coefficient partially linear models.
Their proposed procedure simultaneously selects significant variables in the parametric components and the nonparametric components. 
	\citet{li13} proposed an automatic variable selection procedure using
	smooth-threshold generalized estimating equations, which are based on the generalized  estimating equations (GEE). 
Most of the above-mentioned methods only focused on the fixed dimension $p$. Thus, \citet{wang12} proposed a penalized GEE using a SCAD penalty and proved the asymptotic
properties under the framework of large sample size $n$ and diverging $p$. The important feature of  their method is that the
	consistency of model selection holds even if the working correlation structure is misspecified. However, the methods mentioned above are all based on the GEE.
	When the longitudinal data are contaminated or follow a heavy-tailed distribution, these methods are sensitive to response and/or covariates outliers. 
For example, in a large-scale yeast cell gene expression study reported by \citet{spellman98}, genome-wide mRNA
	levels for $6178$ yeast  open reading frames  that can determine which amino acids will be encoded by a gene were recorded.
	The yeast cell cycle gene expression data cover approximately two cell-cycle periods and were collected  at 7-minute
	intervals for 119 minutes, for a total of 18-time points measured at M/G1-G1-S-G2-M stages. Figure \ref{ycor} reveals that there
	exist strong correlations and the correlation matrix is   never a commonly used exchangeable or autoregressive matrix.
	Furthermore, apart from the strong correlations, we find that there are some outliers in the gene expression data (see Figure \ref{youtlier}).
	Figure \ref{Xoutlier} indicates that abundant influence points occur in the observations of some important transcription factors,
	such as ASH1, MBP1, SWI4, and SWI6, which may lead to biased estimation and prediction.
	Some researchers have proposed various methods to identify important transcription factors (TFs) from a large set of transcription factors that
	are associated with gene expression levels and capture a complex relationship among those factors \citep{luan03,wang07,wang12}.
	Nevertheless, few researches focus on the robustness against outliers on observations and most of them fail to capture the underlying correlation structure within gene expression level on multiple
	observations.

Robust methods are desirable for contaminated data. Therefore, \citet{fan12} proposed robust penalized estimating equations based on Huber's function for linear regression with
longitudinal data, which is  robust against outliers in response, but is sensitive to outliers in covariates. The regulated  parameter in Huber's function is
	directly specified.
\citet{lv15} explored a weighted variable selection method
based on an exponential squared loss \citep{Wang13} and a commonly used working correlation matrix for high dimensional longitudinal data,
and they also provided a data-driven method to select the parameter in the exponential squared loss. These two methods are robust, but their work only looked at a specific case in which the
variable dimension was no larger than the sample size, that is $p < n$.

In this article, we construct robust weighted estimating functions based on Tukey's biweight score equations, which are robust for outliers in response and/ or covariates. 
Different from  \citet{li13} using robust residuals to estimate the correlation parameter, we   propose a novel robust working correlation matrix to capture the correlations, which is more close to the true correlation matrix than the exchangeable and AR(1) correlation matrices, and performs competitively with the true correlation structure in variable selection.
Following \citet{li13} and \citet{Chang18}, we establish robust smooth-threshold estimating equations for parameter estimation and variable selection.
Furthermore, we prove the asymptotic  properties of the proposed method under ``large $n$ and diverging $p$'' setting.
Robust estimating equations using bounded scores and leverage-based weights are  robust against outliers and can reduce the bias when errors
follow a heavy-tailed distribution. The proposed method can be applied to sparse marginal
models under the large $n$ small $p$, large $n$ diverging $p$, and small $n$ large $p$.
%and ``large $n$, fixed $p$''.

The rest of the article is organized as follows: In Section 2.1, we construct a robust estimating equation (RTGEE) for parameter estimation and variable selection.
In Section 2.2, we apply an iterative algorithm to solve the smooth-threshold generalized estimating equations. In Section 2.3, we establish an effective criterion for tuning parameter selection.
In Section 3, we establish the oracle properties of the proposed method. In Section 4, we carry out extensive simulation studies to evaluate the performance of the proposed method. In Section 5, we analyze a yeast cell cycle dataset to illustrate the proposed method. Finally, in Section 6, we draw some conclusions.

\section{Robust smooth-threshold GEE}
Suppose that $Y_{i}=(y_{i1},\ldots,y_{im_i})^{\rm T}$ are
measurements collected at times $(t_{i1},\ldots, t_{im_i})$ for the
$i$th subject, where $i=1,\ldots, n$. Let $X_i=(x_{i1},\ldots,
x_{im_i})$ be the corresponding covariate vector, in which
$x_{ij}=(x_{ij1},\ldots,x_{ijp})^{\rm T}$ is a $p \times 1$ vector. Assume that observations from the
same subject are correlated, and observations from different subjects are independent.
Denote the marginal mean of  $y_{ij}$ by $\mu_{i j}=E\left(y_{i j} | {x}_{i j}\right)=g\left(x_{i j}^{\top}\boldsymbol{\beta}\right)$,
where $g(\cdot)$ is the inverse of the known link function, $\boldsymbol{\beta}=(\beta_1,\ldots,\beta_p)^{\rm T}$ is an unknown parameter vector, and
variance of $y_{ij}$ is
$\operatorname{Var}\left(y_{i j} | x_{i j}\right)=\phi v\left(\mu_{i j}\right)$ with
a variance function $v(\cdot)$ and a scale parameter $\phi$. %Here $g(\cdot)$ is a link function.
Let $\mu_i=(\mu_{i1},\ldots, \mu_{im_i})^{\rm T}$ and ${A}_{i}=\phi \operatorname{diag}\left(v\left(\mu_{i 1}\right), \ldots, v\left(\mu_{i m_{i}}\right)\right)$
be a diagonal matrix. The covariance matrix of $Y_i$ is ${Cov(Y_i)}={A_i}^{1/2}{R_{T}}{A_i}^{1/2}$, where $R_T$ is the true correlation matrix of $Y_i$.
%In generalized estimating equations, working correlation matrix $R(\alpha)$ is a $m_i\times m_i$ symmetric matrix that involves an unknown
%parameter vector $\alpha$. There are some commonly used working correlation structures including independence matrix, autoregressive matrix, and exchangeable matrix.
%For a given working correlation structure, $\alpha$ can be estimated using residual-based moment method \cite{liang86}. If $R(\alpha)$ is the true correlation matrix of $Y_i$'s,
%then $A_i^{1/2}R(\alpha)A_i^{1/2}$ will be equal to the covariance matrix of $Y_i$.

\subsection{Methodology}
We consider a new efficient and robust Tukey's biweight generalized estimating equation (RTGEE) for marginal longitudinal data:
\begin{eqnarray}
%{U}_{n}^{\mathrm{RTGEE}}{(\boldsymbol{\beta})}=\sum_{i=1}^{n}{D}_{i}^{T}{V}_{i}^{-1}{h}_{i}^{b}\left({\mu}_{i}(\boldsymbol{\beta})\right),
U_{n}(\boldsymbol{\beta},\alpha)=\sum_{i=1}^{n} U_{i}(\boldsymbol{\beta})=\sum_{i=1}^{n}{D}_{i}^{T}{V}_{i}^{-1}{h}_{i}^{b}\left({\mu}_{i}(\boldsymbol{\beta})\right)=0,
\end{eqnarray}
where ${D}_{i}=\partial {\mu}_{i} / \partial {\boldsymbol{\beta}}$,
$V_{i}=R_{i}(\alpha) A_{i}^{\frac{1}{2}}$, ${h}_{i}^{b}\left({\mu}_{i}\right)={W}_{i}\left[\widetilde\psi_{b}\left({\mu}_{i}(\boldsymbol{\beta})\right)-C_{i}\left({\mu}_{i}(\boldsymbol{\beta})\right)\right]$
	with $C_{i}\left({\mu}_{i}\right)=E\left[\widetilde\psi_{b}\left({\mu}_{i}(\beta)\right)\right]$,
and $W_{i}$ is a diagonal weight matrix used to downweight the effect of leverage points. One such leverage point, the $j$th element, is
$$
w_{i j}=w\left(x_{i j}\right)=\min \left\{1,\left\{\frac{b_{0}}{\left(x_{i j}-m_{x}\right)^{T} S_{x}^{-1}\left(x_{i j}-m_{x}\right)}\right\}^{\frac{r}{2}}\right\},
$$
where $r \geq 1$, $b_0$ is the 0.95 quantile of the $\chi^2$ distribution with $p$ degrees of freedom, and $m_x$ and $S_x$ are
some robust estimators of the location and scale of $x_{ij}$. % such as the median of $x_{ij}$ and the median absolute deviance (MAD) of $x_{ij}$.
The robust function $\widetilde\psi_{b}\left({\mu}_{i}\right)=\psi_{b}\left({A}_{i}^{-1 / 2}\left({Y}_{i}-{\mu}_{i}\right)\right)$ is given as follows:
$$
\psi_{b}(u)=\left\{\begin{array}{ll}
{u[1-\left(\frac{u}{b}\right)^{2}]^{2}} & {\text { if }|u| \leq b} \\
{0} & {\text { if }|u|>b}
\end{array}\right.,
$$
which is the derivative of Tukey's biweight loss function. %The criterion of choosing optimal parameter $b$ is presented in Section 2.3.

Guided by the idea of \citet{Ueki09}, we select important variables via an efficient and robust smooth-threshold GEE:
\begin{eqnarray}\label{ee}
%\left({I}_{p}-{\Delta}\right) U_{n}^{RTGEE}{(\boldsymbol{\beta}, \alpha)}+\Delta \boldsymbol{\beta}={0},
\left({I}_{p}-{\Delta}\right) U_{n}(\boldsymbol{\beta},\alpha)+\Delta \boldsymbol{\beta}={0},
\end{eqnarray}
where ${I}_{p}$ is the $p$-dimensional identity matrix, and $\Delta=\diag\{\hat{\delta}_{1},\hat{\delta}_{2},\ldots,\hat{\delta}_{p}\}$ is a diagonal matrix,
in which  $\hat{\delta}_{j}=\min \left\{1, \lambda /\left|\hat{\beta}_{j}^{(0)}\right|^{(1+\tau)}\right\}$ with a consistent
estimator $\hat{\beta}_j^{(0)}$ of $\beta_j$.
When $\hat{\delta}_{j}=1$, we shrink
$\hat{\beta}_{j}$ to zero and thus obtain a sparse estimator. The parameter $\tau$ can be selected among (0.5, 1, 2)
according to a suggestion from numerical studies in \citet{zou06}. In simulation studies, we found that  $\tau=1$ is highly effective
for the numerical simulations.

Let $\hat{\boldsymbol{\beta}}$ be a consistent estimator of $\boldsymbol{\beta}$, and let
$\hat e_{i}=\left(\phi A_{i}\right)^{-1 / 2}\left(Y_{i}-\mu_{i}(\hat{\boldsymbol{\beta}})\right)$ be the standardized Pearson residuals.
For a chosen score function $\psi_{b}(\cdot)$, corresponding robust residuals are denoted as ${\psi}_{b}\left({e}_{i}\right)=\left\{\psi_{b}\left(e_{i 1}\right), \ldots, \psi_{b}\left(e_{i m_{i}}\right)\right\}^{\rm T}$.
To solve equation (\ref{ee}), we need to specify the working correlation matrix $R_i(\alpha)$.
Instead of estimating a constant correlation parameter for a specific correlation structure such as exchangeable
	and the first-order autoregressive correlation structures,
	here we estimate the correlation coefficient vector $\alpha$
	via constructing a new unstructured correlation matrix which is more close to the true correlation matrix:
\begin{eqnarray}
R_{u}=\frac{1}{n}\sum_{i=1}^n {\psi_{b}}\left(\hat{e}_{i}\right){\psi_{b}^{\rm T}}\left(\hat{e}_{i}\right).
\end{eqnarray}
%where $\hat{e}_{i}=(\phi A_{i})^{-1 / 2}\left(y_{i}-\mu_{i}(\hat{\beta})\right)$ is the standardized Pearson residuals, and $\hat{\boldsymbol{\beta}}$ is a consistent estimator of $\boldsymbol{\beta}$.
To guarantee the diagonal elements of $R_{u}$ are equal to 1, and the off diagonal elements of $R_u$  belong to $(-1,1)$,
we reconstruct the working correlation matrix $R_u$ and propose the following matrix:
\begin{eqnarray}\label{run}
R_{un}=B_o^{-1/2}R_uB_o^{-1/2},
\end{eqnarray}
where
$B_o=\diag(\sum_{i=1}^n {{\psi_{b}^2}\left(\hat{e}_{i1}\right)}/n,\sum_{i=1}^n {{\psi_{b}^2}\left(\hat{e}_{i2}\right)}/n,\ldots, \sum_{i=1}^n {{\psi_{b}^2}\left(\hat{e}_{im_i}\right)}/n)$.
Accordingly, we assign $R_{un}$ as an estimate of the working correlation matrix $R_i(\alpha)$, in which $\alpha$ is a correlation parameter vector.
Hence, the diagonal elements of $R_{un}$ are equal to $1$, and the off diagonal elements of  $R_{un}$ which are  estimates of the  vector  $\alpha$ always lie in $(-1,1)$ %, that is because
%$$
%(\sum_{i=1}^n {{\psi_{b}}\left(\hat{e}_{iu}\right)} {{\psi_{b}}\left(\hat{e}_{iv}\right)})^2\leq \sum_{i=1}^n {\psi_{b}^2}\left(\hat{e}_{iu}\right)\sum_{i=1}^n {\psi_{b}^2}\left(\hat{e}_{iv}\right)
%$$
according to Cauchy--Schwarz inequality.
%\textcolor{blue}{Accordingly, the correlation parameter estimate of vector $\alpha$ consists of elements of $R_{un}$ in the working correlation matrix $R_i(\alpha)$.}

To obtain the standardized Pearson residuals $\hat e_{i}$, we need to specify the scale parameter $\phi$. Here, we use the robust median absolute deviation to estimate  $\phi$ \citep{wang05}:
\begin{equation}\label{phi}
\hat{\phi}=\left\{1.483 \text { median }\left\{\left|\hat{\eta}_{ij}-\operatorname{median}\left(\hat{\eta}_{ij}\right)\right|\right\}\right\}^{2},
\end{equation}
where $\hat{\eta}_{i j}=A_{i j}^{-1 / 2}\left(y_{i j}-\mu_{i j}(\hat{\beta})\right)$.

\subsection{Algorithm}
To select the important variables and estimate the regression parameters in the marginal models, we follow a Fisher scoring iterative algorithm to implement the procedures as follows:\\
Step 1. Give an initial estimator $\hat{\boldsymbol{\beta}}^{(0)}$, for example, one can use the MM-estimator as an initial value to ensure stability. Let $k=0$.\\
%And for GLMs, for example, in high-dimensional Poisson linear regression, one can choose estimators of log-linear model using glm function in `MASS' package in R.
Step 2. Estimate the scale parameter $\hat \phi$ using (\ref{phi}) with the current estimator $\hat{\boldsymbol{\beta}}^{(k)}$. Compute the working correlation
matrix $\hat{R}_i$ using (\ref{run}), and we get
$${V}_{i}\left\{{\mu}_{i}\left(\hat{\boldsymbol{\beta}}^{(k)}\right), \hat{\phi}^{(k)}\right\}=\hat{R}_{i} \hat{{A}}_{i}^{1 / 2}\left(\hat{\boldsymbol{\beta}}^{(k)}, \hat{\phi}^{(k)}\right).
$$
Step 3. For a given $\lambda$, we update the estimator of $\beta$ via the following iterative formula:
\begin{eqnarray}
\left.\hat{\boldsymbol{\beta}}^{(k+1)}=\hat{\boldsymbol{\beta}}^{(k)}-\left\{\left(\sum_{i=1}^{n} D_{i}^{T} \Omega_{i}\left(\mu_{i}(\boldsymbol{\beta})\right) D_{i}+\hat{G}\right)^{-1}
\left(U_{n}(\boldsymbol{\beta})+
\hat{G}\boldsymbol{\beta} \right)\right\}\right|_{\boldsymbol{\beta}=\hat{\boldsymbol{\beta}}^{(k)}},
\end{eqnarray}
where $\hat{G}=\left(I_p-\hat \Delta\right)^{-1}\hat \Delta$, and ${\Omega}_i\left({\mu}_{i}({\boldsymbol{\beta}})\right)={V}_{i}^{-1}\left({\mu}_{i}({\boldsymbol{\beta}})\right) {\Gamma}_{i}\left({\mu}_{i}({\boldsymbol{\beta}})\right)$, in which
$$
\Gamma_i{(\mu_i{(\boldsymbol{\beta})})}=E\left[\dot{h}^{b}_{i}{(\mu_i(\boldsymbol{\beta}))}\right]
=E\left.\left[\partial{h}^{b}_{i}{(\mu_i(\boldsymbol{\beta}))}/\partial{\mu_i}\right]\right|_{\mu_i=\mu_i(\boldsymbol{\beta})},
$$
Step 4. Repeat Steps 2--3 until the algorithm converges.
Here we set stop condition $||\hat{\boldsymbol{\beta}}^{(k+1)}-\hat{\boldsymbol{\beta}}^{(k)}||^2<\epsilon$, where $\epsilon$ is a small number and
takes a fixed value of $\epsilon=10^{-8}$.

With the given $\lambda$ and $b$, the corresponding parameter estimator of $\boldsymbol\beta$ is denoted as $\hat{\beta}^b_{\lambda}$. %is the estimator of $\beta$.
According to the iterative algorithm mentioned above, we obtain a sandwich formula to estimate the asymptotic covariance matrix of $\hat{\boldsymbol{\beta}}^b_{\lambda}$:
\begin{eqnarray}\label{covm}
\Cov(\hat{\boldsymbol{\beta}}^b_{\lambda})\approx\left[\hat{\mathbf{\Sigma}}_{n}\left({\mu}_{i}(\hat{\boldsymbol{\beta}}^b_{\lambda})\right)\right]^{-1} \hat{\mathbf{H}}_{n}\left({\mu}_{i}(\hat{\boldsymbol{\beta}}^b_{\lambda})\right)\left[\hat{\mathbf{\Sigma}}_{n}\left({\mu}_{i}(\hat{\boldsymbol{\beta}}^b_{\lambda})\right)\right]^{-1},
\end{eqnarray}
where
$$\hat{\mathbf{H}}_{n}\left(\boldsymbol{\mu}_{i}(\hat{\boldsymbol{\beta}}^b_{\lambda})\right)=\sum_{i=1}^{n} {D}_{i}^{T}{V}_{i}^{-1}\left({\mu}_{i}(\hat{\boldsymbol{\beta}}^b_{\lambda})\right)\left[{h}_{i}^{b}\left({\mu}_{i}(\hat{\boldsymbol{\beta}}^b_{\lambda})\right)\left\{{h}_{i}^{b}\left({\mu}_{i}(\hat{\boldsymbol{\beta}}^b_{\lambda})\right)\right\}^{T}\right] {V}_{i}^{-1}\left({\mu}_{i} (\hat{\boldsymbol{\beta}}^b_{\lambda})\right) {D}_{i}^{T},$$
and
$$\hat{\mathbf{\Sigma}}_{n}\left({\mu}_{i}(\hat{\boldsymbol{\beta}}^b_{\lambda})\right)=\sum_{i=1}^{n} {D}_{i}^{T} {V}_{i}^{-1}\left({\mu}_{i}(\hat{\boldsymbol{\beta}}^b_{\lambda})\right) {\Gamma}_{i}\left({\mu}_{i}(\hat{\boldsymbol{\beta}}^b_{\lambda})\right) {D}_{i}.$$
%One can also obtain sparse solutions for smooth-threshold RGEE \cite{fan12} and ERSGEE \cite{lv15} by solving (9)
%under robust estimating equations $U_{n}^{RGEE}(\beta)$ ,${U}_{n}^{\mathrm{ERGEE}}(\beta)$ in the forms of (1) and (3) respectively.

\subsection{Selection of tuning parameters}
To effectively select important variables using the  proposed method, we need to choose proper tuning parameters $b$ and $\lambda$ as mentioned in Section 2.2, which determines the
robustness of the estimator and consistency of variable selection respectively.
For a given $\lambda$, we select the optimal parameter $b$ in $\psi_{b}(u)$  from a series of candidates by minimizing the determinant value of the covariance matrix of $\hat{\boldsymbol{\beta}}^b_{\lambda}$:
\begin{equation}\label{b}
b^{opt}_{\lambda}=\min \limits_{b} \operatorname{det}(\operatorname{Cov}(\hat{\boldsymbol{\beta}}^b_{\lambda}))
\end{equation}
The covariance matrix $\operatorname{Cov}(\hat{\boldsymbol{\beta}}^b_{\lambda})$  can be obtained from (\ref{covm}). In our simulations, we take a series of candidates satisfying asymptotic efficiency higher than $0.7$ compared to the Gaussian distribution, which have been
	listed in Table 2 in \citet{Riani14}.  For the regularization parameter $\lambda$ selection, we adopt the PWD-type criterion proposed by \citet{li13} to choose regularization parameter $\lambda$ for (\ref{ee}):
\begin{eqnarray}
	\mbox{RPWD}_\lambda=\sum\limits_{i=1}^{n}\{h_i^{b_{\lambda}^{opt}}(\mu_i(\hat{\boldsymbol{\beta}}_{\lambda}))\}^TR_i^{-1}(\mu_i(\hat{\boldsymbol{\beta}}_{\lambda}))\{h_i^{b_{\lambda}^{opt}}(\mu_i(\hat{\boldsymbol{\beta}}_{\lambda}))\}+df_\lambda \log(n),
	\end{eqnarray}
	where $\hat{\boldsymbol{\beta}}_{\lambda}$ is the estimator of $\beta$ for a given $\lambda$ and the corresponding optimal $b$ as in (\ref{b}).
Denote $df_\lambda=\sum\limits_{j=1}^{p} 1 (\hat{\delta_j} \neq 1)$ as the number of nonzero elements of the estimators.
% $\hat{\delta_j}=min\{1, \lambda/|\hat{\beta}_j^{(0)}|^{2}\}$,and $\hat{\beta}_j^{(0)}$ is an initial estimator mentioned in Section 2.2.
We choose $\lambda$, which corresponds to the minimizer of $\text{RPWD}_\lambda$, as an optimal value among a series of candidate values with a convergent
	solution $\hat{\boldsymbol{\beta}}^{b}_{\lambda}$ under each $\lambda$ and $b$ values.

\section{Asymptotic properties}
\label{s:properties}
In this section, we will establish large sample properties of the proposed estimator under a ``large $n$, diverging $p$'' framework, which allows $p_{n}$ to diverge to $\infty$ as $n$ increases.
The detailed proof of following Propositions are presented in the Appendix 1 in the Supplementary Information.
%We also establish the theoretical properties under ``large $n$, fixed $p$'', which are presented in the Appendix 1 of Supporting Information.

Let $\boldsymbol{\beta}_0=(\beta_{01},\ldots,\beta_{0p_n})^{\rm T}$ be the true value of $\boldsymbol{\beta}$,
where $\boldsymbol{\beta} \in \mathbf{\Theta}, \mathbf{\Theta} \subseteq \mathbb{R}^{p_n}$ is
a bounded $p_n$-dimensional vector. Without loss of generality, we denote $\boldsymbol{\beta}_{0}=\left(\boldsymbol{\beta}_{01}^{T}, \boldsymbol{\beta}_{02}^{T}\right)^{T}$,
where $\boldsymbol{\beta}_{02}=\mathbf{0}$, and the elements of $\boldsymbol{\beta}_{01}$ are assumed to be nonzero in the dimension of $s_n$, which can also diverge with $n$.
We partition $\boldsymbol{\beta}_0$ into active (nonzero) coefficient sets $\mathscr{A}_{0}=\left\{j: \beta_{0 j} \neq 0\right\}$ with $|\mathscr{A}_{0}|=s_n$
and inactive (zero) coefficient sets $\mathscr{A}_{0}^{c}=\left\{j: \beta_{0 j}=0\right\}$.
We define the active set $\mathscr{A}=\left\{j: \hat{\delta}_{j} \neq 1\right\}$ as the set of indices of nonzero estimated coefficients.
Under the following conditions, we present the consistency of the proposed estimator.

C1. Assume $x_{i j}$ for $1 \leqslant i \leqslant n$ and $1 \leqslant j \leqslant m_i$ satisfy $\sup _{i, j}\left\|x_{i j}\right\|=O\left(\sqrt{p_{n}}\right)$.
%are uniformly bounded and that the fourth moments of $y_{ij}$ exists.
%Also, for each $i$, ${m_i}$ is a bounded sequence of positive integers.

C2. The unknown parameter $\boldsymbol{\beta}$ belongs to a compact subset $\mathbf{\Theta} \subseteq \mathbb{R}^{p_n}$, and the true parameter value
	$\boldsymbol{\beta}_0$ lies in the interior of $\mathbf{\Theta}$.
	Furthermore, we assume that the estimator of the correlation parameter vector $\hat{\alpha}$ is $\sqrt{p_n/n}$-consistent given $\boldsymbol \beta$ and $\phi$ for some $\alpha$, that is,
	$\left\|\hat\alpha-\alpha\right\|=O_{p}\left(\sqrt{p_{n} / n}\right)$, %and $\hat \phi$ is $\sqrt{n}$-consistent given $\boldsymbol \beta$,
	and $|\partial \hat{\alpha}(\boldsymbol{\beta}, \phi) / \partial \phi| \leq H({Y}, \boldsymbol{\beta})$, where $H(\cdot,\cdot)$ is a bounded function for samples $Y$ and $\boldsymbol \beta$.

C3. Denote $\mathbf{X}_{i}^{h,b}=\mathbf{X}_{i}^{T}h_{0, i}^{b}\left({e}_{i}\right)$.
	There exists finite positive constants $c_{1} \leq c_{2}$ such that $\forall 1 \leq j \leq m_i$:
	$$
	c_{1} \leq \lambda_{\min }\left(n^{-1} \sum_{i=1}^{n}\mathbf{X}_{i}^{h,b}\left\{\mathbf{X}_{i}^{h,b}\right\}^{T}\right) \leq \lambda_{\max }\left(n^{-1} \sum_{i=1}^{n} \mathbf{X}_{i}^{h,b}\left\{\mathbf{X}_{i}^{h,b}\right\}^{T}\right) \leq c_{2},
	$$
	where $h_{0, i}^{b}\left({e}_{i}\right)$ centers ${Y}_{i}$ by its true mean ${\mu}_{0, i}$ with ${\mu}_{0, i}={\mu}_{i}(\boldsymbol{\beta}_0)$.

C4. $\sup _{i \geq 1} E\left\|h_{0, i}^{b}\left({e}_{i}\right)\right\|^{2+\delta}<\infty $  for some $\delta>0$, and $0<\sup_{i} \|E h_{0, i}^{b}\left({e}_{i}\right)\left({h}_{0, i}^{b}\left({e}_{i}\right)\right)^{T}\|<\infty$,
	where ${h}_{i}^{b}\left({e}_{i}\right)$ centers by ${\mu}_{i}={\mu}_{i}(\boldsymbol{\beta})$.

C5. There exists a positive constant $c$ such that $0<c \leq \inf _{i, j} v\left(\mu_{i j}\right) \leq \sup _{i, j} v\left(\mu_{i j}\right)<\infty$.
The functions $C_{i j}\left(\mu_{i j}\right)=E\left[\psi_{b}\left(A_{i j}^{-1 / 2}\left(y_{i j}-\mu_{i j}\right)\right)\right]$,  $v(\cdot)$ and $ g(\cdot)$ have bounded second derivatives.
The function $\psi_{b}(\cdot)$ is piecewise twice differentiable, and the second derivatives are bounded.

C6. Assume that $E\left\|U_{n}\left(\boldsymbol{\beta}_{0}\right)\right\|^{2}<\infty$ and there exists $\delta>0$ such that
$$
\lim _{n \rightarrow \infty} \frac{\sum_{i=1}^{n} E\left\|U_{i}\left(\boldsymbol{\beta}_{0}\right)\right\|^{2+\delta}}{\left(E\left\|U_{n}\left(\boldsymbol{\beta}_{0}\right)\right\|^{2}\right)^{1+\delta/2}}=0.
$$

C7. Matrix
$$
\boldsymbol{\Sigma}=\lim _{n \rightarrow \infty} n^{-1} \sum_{i=1}^{n}\left[{D}_{0, i}^{T} {V}_{0, i}^{-1} {\Gamma}_{0, i}\left({\mu}_{i}\left(\boldsymbol{\beta}_{0}\right)\right) {D}_{0, i}\right]
$$
is positive definite.
Matrix
$$
\boldsymbol{B}=\lim _{n \rightarrow \infty} n^{-1} \sum_{i=1}^{n} {D}_{0, i}^{T} {V}_{0, i}^{-1} \operatorname{cov}\left({h}_{i}^{b}\left({\mu}_{i}\left(\boldsymbol{\beta}_{0}\right)\right)\right)\left({V}_{0, i}^{-1}\right)^{T} {D}_{0, i}
$$
is also positive definite.

C8. For any positive $\lambda$, $\tau$, $p_n$, and $s_n$,
${(p_n/n)}^{(1+\tau) / 2}{\lambda}^{-1} \rightarrow 0$, $n^{-1/2}{\lambda}^2=o(1)$, and $s_{n} n^{-1/2}=o(1)$, such as $s_{n}=O(n^{1/3})$.

{\bf Remark} Condition C1 is a common assumption in the M-estimator with diverging dimension \citep{portnoy85}, and it holds almost surely under some weak
	moment conditions for $x_{ij}$ from spherically symmetric distributions. Condition C2 is established to ensure the $\sqrt{p_n/n}$-consistency
	of $\hat R_i(\alpha)$ in Section 2.1, which can be verified using similar analysis in \citet{he05}. Taking the spirit of Lemma 3.7 in \citet{wang11},
	we set condition C3, which is especially useful when establishing the asymptotic normality in Proposition 2. Similar to \citet{lv15},
	conditions C4--C5 can be easily checked under bounded score function $\psi_{b}(\cdot)$, and they are usually combined with assumptions C6--C7, which are also necessary
	for the central limit theory and hold in most cases. Condition C8 is established for exploring convergence rate and asymptotic properties, which controls the
	order of diverging number $p_n$ and $s_n$ precisely, and we point out a series theoretical values for regularization parameter $\lambda$. Note that
	a preliminary $\sqrt{p_n/n}$-consistent estimator $\boldsymbol{\beta}_0$ is needed in both Proposition 1 and Proposition 2, which can be obtained
	by solving the generalized estimating equations under independent working correlation structure as in Example 1 in \citet{wang11} when $p_n\rightarrow \infty$.

{\bf Proposition 1} Suppose the regularity conditions C1--C8 hold, then we have
$$
\left\|\hat{\boldsymbol{\beta}}_{\lambda, \tau}-\boldsymbol{\beta}_{0}\right\|=O_{p}\left(\sqrt{p_{n}/n}\right).
$$

{\bf Proposition 2} Under conditions C1--C8, and if $n^{-1} p_{n}^{3}=o(1)$, as $n \rightarrow \infty$, we have\\
	(1) variable selection consistency, $P\left(\mathscr{A}=\mathscr{A}_{0}\right) \rightarrow 1$;\\
	(2) asymptotic normality: $\forall \boldsymbol{\alpha}_{n} \in R^{s_{n}}$ such that $\left\|\boldsymbol{\alpha}_{n}\right\|=1$,
	$$
	\sqrt{n}\boldsymbol{\alpha}_{n}^{T}\mathbf{B}_{\mathscr{A}_{0}}^{-1/2} \boldsymbol{\Sigma}_{\mathscr{A}_{0}}\left(\hat{\boldsymbol{\beta}}_{\lambda, \tau, \mathscr{A}}-\boldsymbol{\beta}_{\mathscr{A}_{0}}\right)\stackrel{d}{\rightarrow} N\left(0, 1 \right),
	$$
	where $\boldsymbol{\Sigma}_{\mathscr{A}_{0}}$ and $\mathbf{B}_{\mathscr{A}_{0}}$ are the first $s_n \times s_n$ submatrices of $\boldsymbol{\Sigma}$ and $\boldsymbol{B}$.

Proposition 1 implies that our proposed estimator can achieve $\sqrt{p_n/n}$-consistency. Proposition 2 shows that such consistent estimators possess
the sparsity property and oracle property \citep{fan01} when we choose proper $\lambda$ and $\tau$. With a probability approaching 1, our proposed method can correctly select
the nonzero coefficients and estimate them as efficiently as if we know the correct submodel in advance. %\textcolor{blue}{The proofs of Proposition 1 and Proposition 2 are given in the Appendix 1.}
%The proofs of Proposition 1 and Proposition 2 are given as supplementary materials and are available at the Statistics in Medicine website.

\section{Simulation studies}
We conduct simulation studies to assess the performance of the proposed RTGEE method,
the smooth-threshold generalized estimating equation (SGEE) proposed by \citet{li13},
the robust smooth-threshold generalized estimating equation (RSGEE) corresponding to Huber's score function, and the efficient and
robust generalized estimating equation (ERSGEE) proposed by \citet{lv15}
for continuous normal data and heavy-tailed data under setups $p<n$, large $n$ and diverging $p$, and $p>n$.
%for continuous data and count data under setups ``$p<n$", large $n$ and diverging $p$, and $p>n$.

For each procedure, the true correlation structure of the response is exchangeable (EXC) with the correlation coefficient $\alpha=0.7$.
For each setup in the simulations, we generate $100$ datasets and apply the iterative algorithm mentioned in Section 2.2 to estimate $\boldsymbol{\beta}$ and select important variables
at the same time. Furthermore, we also consider the situation in which the true correlation structure is AR(1) with correlation parameter $\alpha=0.7$ for the continuous data. The simulation results show similar
patterns and are presented in Tables 1--6 in the supplementary materials. Finally, we also consider the count data, and the results are listed in Tables 7--8 in the supplementary materials.

\begin{figure}
	\begin{center}
		\includegraphics[width=0.5\textwidth]{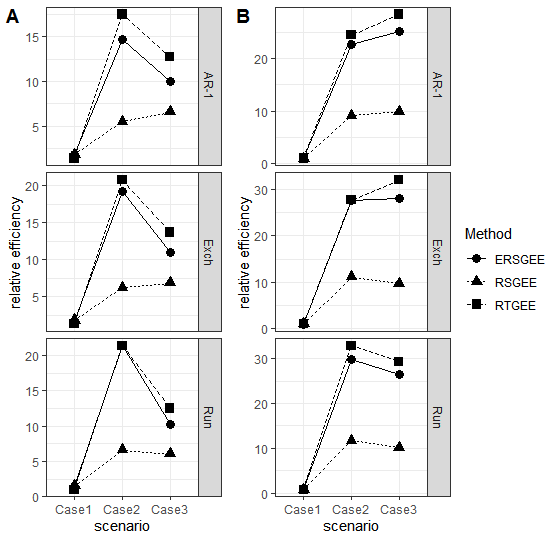}
	\end{center}
	\caption{Comparison of RSGEE, ERSGEE, and RTGEE on relative efficiency (compared to SGEE) with three different working correlation matrices under three contamination scenarios. The left figure A represents simulation \Rmnum{1} for heavy-tailed data, and the right
		figure B represents simulation \Rmnum{1} for the normal data.\label{fig_tab1}}
\end{figure}

\begin{figure}
	\begin{center}
		\includegraphics[width=0.5\textwidth]{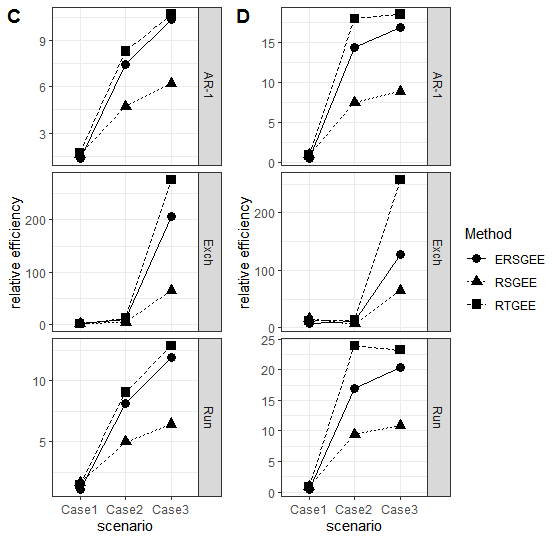}
	\end{center}
	\caption{Comparison of RSGEE, ERSGEE, and RTGEE on relative efficiency (compared to SGEE) with three different working correlation matrices under three contamination scenarios. The left figure C represents simulation \Rmnum{2} for heavy-tailed data, and the right
		figure D represents simulation \Rmnum{2} for the normal data.\label{fig_tab2}}
\end{figure}

\begin{figure}
	\begin{center}
		\includegraphics[width=0.5\textwidth]{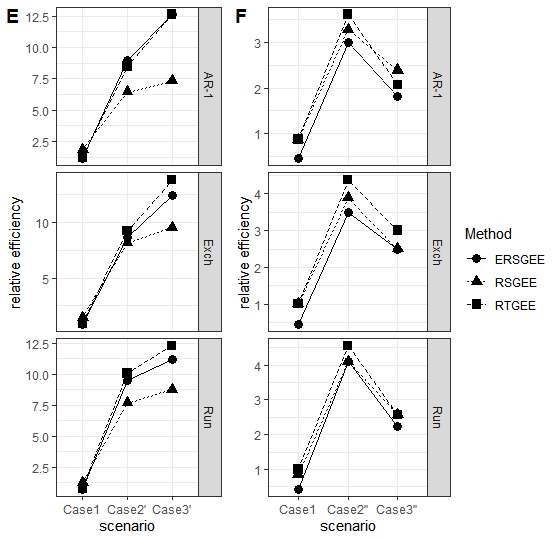}
	\end{center}
	\caption{Comparison of RSGEE, ERSGEE, and RTGEE on relative efficiency (compared to SGEE) with three different working correlation matrices under three contamination scenarios. The left figure E represents simulation \Rmnum{3} for the heavy-tailed data, and the right
		figure F represents simulation \Rmnum{3} for the normal data.\label{fig_tab3}}
\end{figure}

We compare these four methods under three working correlation matrices (EXC, AR(1), $R_{un}$) according to the following terms:
the average number of correctly identified insignificant variables (C), the average number of incorrectly identified significant variables (IC), the correctly fitted odds (CF, the odd of
identifying both significant variables and insignificant variables correctly over 100 simulations),
the biases of estimators, the standard deviance (SD) of estimators,
the proportion of estimators fall into the 95\% confidence interval (CI),
the average of mean squared prediction error (AMSPE), the median of mean squared prediction error (MMSPE),
where $\text{MSPE}=n^{-1} \sum_{i=1}^{n}\left(\hat{y}_{i}-y_{i}\right)^{2}$, and the average mean square error (AMSE), which is the average of $\left\|\hat{\beta}-\beta_{0}\right\|^{2}$ over 100 simulations. To demonstrate the efficiency of estimators, we compare the relative efficiency among three robust methods in Figures \ref{fig_tab1}--\ref{fig_tab3}, which is defined as the ratio of the AMSE for SGEE to the AMSE
	for each robust method, from which a higher value represents higher efficiency.
	We present partial results in Tables 1--6 and more details can be found in Tables 9--14 in Appendix 2 in the supplementary information.

%\subsection{Continuous data}
%\label{ss:simulations}
\subsection{Heavy-tailed continuous data}
We generate the continuous data from the following model:
\begin{eqnarray}\label{simu1}
y_{ij}=x_{ij1}\beta_{1}+x_{ij2}\beta_{2}+\cdots+x_{ijp}\beta_{p}+\epsilon_{ij}, ~i=1,\ldots,n , ~j=1,\ldots,m_i.
\end{eqnarray}
Without loss of generality, we consider the balanced data with $m_i=10$ for $i=1,\ldots,n$.
Covariates $x_{ij}=(x_{ij1},\ldots,x_{ijp})^{\rm T}$ follow a multivariate normal distribution with a mean of zero and the correlation between the $k$th and $l$th component of $x_{ij}$ being
$0.5^{|l-k|}$. The random error vectors $\epsilon_i=(\epsilon_{i1},\ldots,\epsilon_{i10})^{\rm T}$ are generated from a multivariate
Student's $t$-distribution with three degrees of freedom $T_3(0,R(\alpha))$.
The true coefficients are assumed to be $\boldsymbol{\beta}=(0.7,0.7,-0.4,0,\ldots,0)^T$ with nonzero coefficients $d=3$ and $p-d$ coefficients being zero.

\Rmnum{1}.  We first test performance when $p=20$ and $n=100$ under the following scenarios:

Case $1$: There is no contamination on the dataset.

Case $2$: We randomly add 20\% $y$-outliers following $N(10,1)$ on $y_{ij}$.%, and we denote new response sets as $Y^{*}$.\\

Case $3$: We randomly add 10\% $x$-outliers on $x_{ij1}$, following a Student's $t$-distribution with three degrees of freedom. %, thus generating a new design matrix $X^{*}$.
Meanwhile, we change response variables in the same way as Case $2$.% There are both $x$-outliers and $y$-outliers for new data sets $(X^{*},Y^{*})$.

\Rmnum{2}.  Next we consider ``large $n$ and diverging $p$".

The true coefficients are set as $\boldsymbol{\beta}=(0.7,0.7,-0.4,0.7,0.7,-0.4,\ldots,\mathbf{0}_{p_n-s_n})$ with $p_n=[4n^{2/5}]-5$ and the scale of nonzero coefficients $s_n=[p_n/5]$ for $n=200$, where $[s]$ denotes the largest positive integer value not greater than $s$. In addition, the observation times $m_i$ are randomly generated from 2 to 5. The other settings are same as those in $p<n$.

\Rmnum{3}. We set $p=300$ and $n=100$, and the true coefficients vector $\beta=(0.7,0.7,-0.4,0,\ldots,0)^T$ is a $p$-dimensional
vector with only three nonzero components. Other conditions are the same as $p<n$.
To ensure the stability of simulations, we decrease the proportions of outliers as follows:

Case $2'$: We randomly add 10\% $y$-outliers following N(10,1) on $y_{ij}$.

Case $3'$: We randomly add 10\% $x$-outliers on $x_{ij1}$ following a Student's t distribution with three degrees of freedom, and we randomly add 10\% $y$-outliers,
similar to Case $2'$.

The simulation results for \Rmnum{1}, \Rmnum{2}, and \Rmnum{3} are presented in Tables \ref{table1-1}--\ref{table3-1}, respectively.
From Table \ref{table1-1}, it is evident that non-robust SGEE has manifest shortcomings in variable selection compared with the other three robust methods according to the value of CF
even in the no contamination case.
When there are outliers in data sets, the defect of SGEE shows more clearly no matter the variable selection or coefficient estimation. In contrast, the three
robust methods (RSGEE, ERSGEE, and RTGEE) perform well, even in a misspecified correlation structure. However, when adding outliers to the response variables, ERSGEE and our proposed
method, RTGEE, perform better in terms of IC and CF than SGEE and RSGEE. RTGEE is more competitive with ERSGEE in both coefficients estimation and variable selection when
adding $x$-outliers and $y$-outliers simultaneously.
RTGEE has a smaller estimation error (MMSPE) and higher CF than the other three methods.
Furthermore, we find that, as a type of misspecified correlation structure, the results under $R_{un}$ are superior over AR(1) and competitive with true correlation structure EXC, especially when
there are outliers in the dataset. The $R_{un}$ boosts the performance of non-robust SGEE in variable selection and significantly decreases prediction and estimation error compared to AR(1).
Figure \ref{fig_tab1} depicts the relative efficiency for RSGEE, ERSGEE, and RTGEE under \Rmnum{1} settings. The left plot A shows that our proposed estimator has higher relatively efficiency
	than the other two robust methods for contaminated heavy-tailed data, and it is more obvious when there are both $x$-outliers and $y$-outliers.

When $p$ is diverging, the results in Table \ref{table2-1} indicate that the proposed method is comparable with RSGEE and ERSGEE in Case $1$ and Case $2$ and
performs better than RSGEE and ERSGEE when the covariates have outliers. In Case $3$, our proposed method performs superiorly over the other methods regardless
of the estimation or variable selection, which confirms that our proposed method has superiority under diverging $p$.
We notice that the CI of our proposed estimator always fly floats around 95\%, even with contamination, which implies the asymptotic normality of our proposed estimator
	and gives a numerical validation of Proposition 2 established in Section 3.
It is appealing to us that almost all the methods perform better under our proposed working correlation structure $R_{un}$,
even better than estimated true correlation structure regardless of whether there are outliers. The proposed method outperforms when the data are contaminated (see Figure \ref{fig_tab2}). 

From Table \ref{table3-1} and plot E in Figure \ref{fig_tab3}, we can see that ERSGEE and RTGEE show superiority in variable selection compared to SGEE and
	RSGEE when there are no outliers. When outliers are added, our proposed method is superior to ERSGEE with a lower MMSPE and higher relatively efficiency,
	implying RTGEE can keep robustness against outliers even under $p>n$ and misspecified working correlation structure.

\subsection{Continuous normal data}
We generate response variable according to model (\ref{simu1}), and
the covariates are generated in the same way as in Section 4.1.
The random error vectors $\epsilon_i=(\epsilon_{i1},\ldots,\epsilon_{i10})$ are generated from a $N_{10}(0,R(\alpha))$ with the correlation coefficient $\alpha=0.7$.
Other settings are the same as those in Section 4.1, except that when testing the performance of the foregoing methods in a sparse model under $p>n$, we
decrease the proportions of outliers as follows:

Case $2''$: We randomly convert 10\% of $y_{i j}$ into $y_{i j}+5$.

Case $3''$: We artificially add 5\% $x$-outliers on $x_{ij1}$ following a $t(3)$ distribution as well as $y$-outliers
which are same with Case $2''$.

The corresponding results are listed in Tables \ref{table4-1}--\ref{table6-1}. From Table \ref{table4-1}, when there are no outliers added on response and/or covariates, SGEE can perform
well as expected, whereas, it is inferior to robust methods no matter in parameter estimation or variable selection in contaminated datasets. It is noticeable that
our proposed unstructured working correlation matrix $R_{un}$ performs competitively with the true correlation matrix, and it is superior to wrongly assigned working correlation
matrix (eg. AR(1)), especially when there are outliers on response and/or covariates.

In Table \ref{table5-1}, when $p$ is diverging, SGEE can not perform as well as robust methods even if there are no outliers added. Among robust methods, ERSGEE and RTGEE perform
	similarly well and they are superior to RSGEE when there are outliers in datasets. Furthermore, RTGEE has significant advantages with higher relative efficiency than
	ERSGEE, especially under the wrongly assigned working correlation structure, according to plot D in Figure \ref{fig_tab2}. The reasonable performance of our proposed estimators
	in CI verifies the oracle properties again.

When $p>n$, the results in Table \ref{table6-1} show that SGEE is affected by outliers more obviously in variable selection compared with robust methods. Although RTGEE and ERSGEE perform
relatively similar well, where they are superior to RSGEE greatly when outliers are added on covariates, RTGEE performs better in terms of estimation accuracy
with lower MMSPE.

\section{Real data analysis}
\label{s:real data analysis}
The cell cycle is one of the most important processes for cell growth, DNA replication, chromosome segregation, and daughter cells' division.
	%Therefore, identification of cell cycle regulated genes has become greatly important.
	Investigating the functions of gene expression during the cell cycle process can give an insight into how the cell cycle affects biological
	processes and cell cycle regulation. Transcription factors (TFs) are a critical part of the cell cycle process, where they have been
	shown to influence gene expression by regulating the flow of genetic information from DNA to mRNA during the cell cycle process.
	We are interested in selecting important TFs from a large set of candidates that are associated with yeast gene expression levels.

We apply the proposed RTGEE method to analyze the yeast cell cycle gene expression dataset, which was  mentioned in Section 1. %collected by \cite{Spellman98}
%The dataset includes genome-wide mRNA levels of $6178$ yeast ORFs covering approximately two cell-cycle periods.%, taking measurements at 7-minute intervals for $119$ minutes,
%for a total of 18-time points measured at M/G1-G1-S-G2-M stages, which was also studied by
%%This dataset has been pre-studied by many researchers. For instance, \cite{luan03} introduced a mixed-effects model using B-splines to
%%account for time dependency of the gene expression measurements over time and the noisy nature of the microarray data. \cite{wang07} developed a functional
%%response model with varying coefficients and a group smoothly clipped absolute deviation (SCAD) regression procedure for selecting the transcriptional factors (TFs)
%%involved in gene regulation during a given biological process with varying coefficients. Similarly, \cite{wang08} applied nonparametric time varying-coefficient model
%%to identify the TFs that might be related to the expression patterns of these 297 cell-cycle cregulated genes.
%Wang et al. \cite{wang12}. %applied the penalized GEE to analyze a subset of 297 cell-cycle-regularized genes to each of the five stages using marginal linear model,
%and selected TFs in terms of both numbers and the specific TFs under different correlation structures.
Our  investigation indicates that log-transformed gene expression levels and observations of TFs contain many underlying outliers, thus it is worthwhile to reanalyze the
yeast cell cycle via robust procedures.
%To better understand the phenomenon underlying cell-cycle process, it is important to identify transcription factors (TFs) that regulate the gene expression levels of cell cycle-regulated genes.
In this section, we apply SGEE, RSGEE, ERSGEE, and RTGEE to the dataset of the G1 stage in a yeast cell cycle with $1132$ observations ($283$ cell-cycled-regularized genes observed over 4-time points).
The dataset is available in R package PGEE.

\begin{figure}
	\begin{center}
		\includegraphics[width=0.5\textwidth]{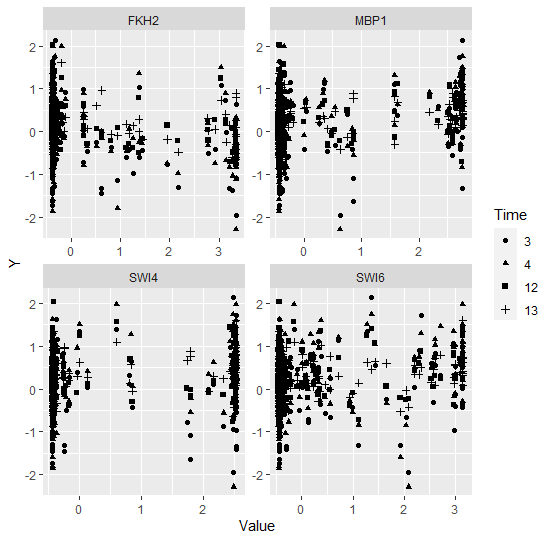}
	\end{center}
	\caption{The scatter plots of gene expression level versus four important TFs: FKH2, MBP1, SWI4, and SWI6. Here ``Y'' represents the log-transformed gene expression level.\label{XYplot}}
\end{figure}

The scatter plot in Figure \ref{XYplot} depicts the complicated functional relationship among gene expression level and TFs, which is highly dependent on varying time, hence we consider
following model, which is the same as \citet{wang12},
$$
y_{i j}=\beta_{0}+\beta_{1} t_{i j}+\sum_{k=1}^{96} \beta_{k} x_{i k}+\epsilon_{i j},~~ i=1,\ldots,283,~~ j=1,\ldots,4,
$$
where the response variable $y_{ij}$ is the log-transformed gene expression level of gene $i$ measured at time point $j$, the covariates $x_{ik}$ are the matching score of the
binding probability of the $k$th transcription factor on the promoter region of the $i$th gene for $k=1,\ldots,96$, and $t_{ij}$ represents the time points.
We consider three correlation structures: EXC, AR(1), and $R_{un}$ for $\epsilon_{i j}$. Table \ref{table7} summarizes the selected numbers of TFs and the mean squared error
for cross validation procedures ($\mathrm{MSE}_{\mathrm{CV}}$) to assess the goodness of fit:
$$
\text{MSE}_{C V}=\frac{1}{n} \sum_{i=1}^{n}\left\|Y_{i}-X_{i} \hat{\beta}_{(-i)}\right\|^{2},
$$
where $\hat{\beta}_{(-i)}$ is the estimator obtained based on the data excluding the $i$th subject. The parameter estimates of the selected TFs are also given in Table \ref{table7}.
%The parameter estimates of selected TFs are given in Table $4^{*}$ in the supplementary.

The results indicate that the robust methods (RSGEE, ERSGEE, and RTGEE) select 25 TFs, while non-robust method (SGEE) select more TFs, from which, we find that
	significant TFs such as MBP1, SWI4, and SWI6 are selected by both robust and non-robust methods, which have been reported to function during the G1 stage in \citet{simon01}.
	In addition, TFs such as FKH2, GAT3, GCR2, NDD1, SRD1, STB1 are commonly selected by all of the methods, and they are also confirmed in \citet{wang12}.
	ABF1 is selected by SGEE in \citet{wang12}, but not by the robust methods. Nevertheless, the transcription factor YAP5 noted as an important
	factor in \citet{ban00} is consistently selected by all of the methods considered in our research, but not selected by \citet{wang12}.
	Similarly, TFs such as MET31 and GCR1 selected by our robust methods have been verified in \citet{tsai05} and \citet{song14} respectively, while
	not been found in \citet{wang12}.

In particular, Table \ref{table7} presents the mean squared error ($\text{MSE}_{\text{C V}}$) and also gives the running time of procedures. 
For this dataset analysis, our proposed
method, RTGEE, performs better than other robust methods with the lower mean squared error under EXC and AR(1), and ERSGEE performs better than RSGEE.
All of the methods using $R_{un}$ can be more competitive than other working correlation structures, though it can be more time-consuming.
	The longer running time of ERSGEE and our procedure RTGEE than other methods is associated with the fact that a wide range of tuning parameters is considered as in Section 2.3.
	To the best of our experience over abundant simulations, we recommend $b=7.0414$ for RTGEE in this yeast data analysis, which can also lead to a sufficient variable selection
	while saving a lot of time.

\section{Conclusions}
This article develops a robust automatic variable selection procedure, RTGEE, in the longitudinal marginal models by utilizing the robustness of Tukey's Biweight criterion.
A new robust working correlation structure is proposed for taking account of the correlations, which is competitive with other misspecified working correlation structures.
According to our simulation results, achieving high effectiveness and consistency in robust variable selection, the proposed working correlation can be a substitute for the
true correlation structure. The correlation parameters in the proposed correlation matrix depend on the number of the repeated measurements $m_i$. When $m_i$ is  large compare with the sample size, the accuracy of the correlation matrix estimation will decrease, and the computation will increase. Hence, the number of the repeated measurements cannot be too large. If  $m=\max_i\{m_i\}$ is diverging or large than the sample size,
	the computation and the theorems need to be restructured, which will be studied in future work. 
We apply smooth-threshold estimating equations to select the important variables. This approach is conceptually simple, easy to implement, and does not need penalty functions.
Furthermore, this approach eliminates the irrelevant parameters by shrinking them to zero and simultaneously estimates the nonzero coefficients.
Previous researchers have proposed similar robust smooth-threshold estimating equations \citep{fan12, lv15}. Nevertheless, the robustness of our method is still competitive
regardless of whether the conditions are regular or there are more severe setting conditions. From our numerical studies, we can conclude that our proposed method is robust
for both response variables and covariates in longitudinal marginal models. It is especially competitive under a heavy tail distribution, and it has broad prospects when the
dimension of covariates is larger than the sample size.

Robust variable selection for ultrahigh-dimensional data is gaining more traction in the biomedical area, and in future research, our proposed method can be extended to cases
where the dimension of covariates is in the exponential order of the sample size.
However, some guiding theoretical research for parameter selection criterion needs to be conducted when applying our procedure to ultrahigh-dimensional data.

\section*{Acknowledgments}

The authors thank the Associate Editor and referees for their constructive comments.  
This research was supported by the Science Foundation of China (No.11871390), and the Natural Science Basic Research Plan in Shaanxi Province of China (No.2018JQ1006), the Australian Research Council Discovery Project (DP160104292).

\begin{table}[H]
	\centering
	\caption{Correlated continuous data for $n>p$ ($p=20$ and $n=100$) with $\epsilon_{ij}$ following a t(3) distribution: Comparison of SGEE, RSGEE, ERSGEE, and the proposed method RTGEE with
		three different working correlation matrices (exchangeable, AR(1) and unstructured).}\label{table1-1}
	%	\noindent{\bf Table 1: Continuous data for $n>p$}\\
	%	{\small We consider linear model with $\epsilon_{i}$
	%		following a multivariate distribution.}\label{tab1}  %heavy tail t(3) distribution
	%\small
	%	\resizebox{\textwidth}{!}{%
	\begin{tabular}{clrcccccccc}
		\hline			
		\multicolumn{3}{c}{ }&
		\multicolumn{1}{c}{$\beta_1$} &
		\multicolumn{1}{c}{$\beta_2$} &
		\multicolumn{1}{c}{$\beta_3$} &
		\multicolumn{1}{c}{ }&
		\multicolumn{2}{c}{No.of Zeros}\\
		\cline{4-6}\cline{8-9}
		Scenario & $R$ & Method & CI   & CI& CI & MMSPE  & C & IC &CF\\
		\hline
		Case 1 &EXC & SGEE &0.96 &0.94 &0.95 & 0.0029 & 16.65 & 0.05 & 0.74 \\
		& & RSGEE          &0.95 &0.94 &0.95 & 0.0016 & 16.91 & 0.00 & 0.92 \\
		& & ERSGEE         &0.93 &0.96 &0.93 & 0.0020 & 16.96 & 0.00 & 0.96 \\
		& & RTGEE          &0.96 &0.95 &0.95 & 0.0022 & 17.00 & 0.00 & 1.00 \\
		&   AR(1) & SGEE   &0.95 &0.94 &0.96 & 0.0042 & 16.42 & 0.02 & 0.60 \\
		& & RSGEE          &0.96 &0.95 &0.98 & 0.0023 & 16.88 & 0.02 & 0.90 \\
		& & ERSGEE         &0.95 &0.94 &0.91 & 0.0023 & 16.95 & 0.00 & 0.95 \\
		& & RTGEE          &0.94 &0.96 &0.95 & 0.0028 & 17.00 & 0.00 & 1.00 \\
		&  $R_{un}$  & SGEE  &0.96 &0.95 &0.95 & 0.0025  & 16.55 & 0.05 & 0.70 \\
		& & RSGEE            &0.95 &0.95 &0.96 & 0.0017 & 16.92 & 0.00 & 0.93 \\
		& & ERSGEE         &0.95 &0.94 &0.94 & 0.0017  & 16.96 & 0.00 & 0.96 \\
		& & RTGEE          &0.95 &0.93 &0.97 & 0.0029  & 17.00 & 0.00 & 1.00 \\
		\hline
		Case 2 &EXC & SGEE & 0.95 & 0.93 & 0.92 & 0.0740 & 15.37 & 0.06 & 0.33 \\
		& & RSGEE          & 0.97 & 0.95 & 0.97 & 0.0078 & 16.89 & 0.01 & 0.90 \\
		& & ERSGEE         & 0.93 & 0.95 & 0.95 & 0.0028 & 16.96 & 0.00 & 0.96 \\
		& & RTGEE          & 0.92 & 0.98 & 0.96 & 0.0021 & 17.00 & 0.00 & 1.00 \\
		&   AR(1) & SGEE   & 0.95 & 0.94 & 0.97 & 0.0807 & 15.06 & 0.00 & 0.27 \\
		& & RSGEE          & 0.96 & 0.95 & 0.98 & 0.0095 & 16.90 & 0.01 & 0.91 \\
		& & ERSGEE         & 0.96 & 0.96 & 0.95 & 0.0036 & 16.96 & 0.00 & 0.96 \\
		& &RTGEE           & 0.92 & 0.97 & 0.95 & 0.0030 & 17.00 & 0.00 & 1.00 \\
		&  $R_{un}$ & SGEE & 0.92 & 0.91 & 0.93 & 0.0768 & 15.63 & 0.03 & 0.38 \\
		& & RSGEE          & 0.96 & 0.95 & 0.98 & 0.0079 & 16.90 & 0.02 & 0.90 \\
		& & ERSGEE         & 0.94 & 0.96 & 0.96 & 0.0026 & 16.93 & 0.00 & 0.94 \\
		& & RTGEE          & 0.94 & 0.97 & 0.99 & 0.0023 & 17.00 & 0.01 & 0.99 \\
		\hline
		Case 3 &EXC & SGEE  & 0.95& 0.95& 0.92 & 8.3204 & 15.94 & 0.05 & 0.42 \\
		& & RSGEE           & 0.90 & 0.96 & 0.96 & 0.0084 & 16.90 & 0.04 & 0.88 \\
		& & ERSGEE & 0.94  & 0.95 & 0.96 & 0.0048 & 16.90 & 0.00 & 0.90 \\
		& & RTGEE  & 0.96  & 0.93 & 0.94 & 0.0027 & 17.00 & 0.06 & 0.94 \\
		&   AR(1)  & SGEE & 0.94 & 0.92 & 0.93 & 0.0728 & 15.86 & 0.03 & 0.45 \\
		& &  RSGEE & 0.93  & 0.96 & 0.95 & 0.0089 & 16.90 & 0.05 & 0.88 \\
		& & ERSGEE & 0.94  & 0.95 & 0.97 & 0.0067 & 16.90 & 0.00 & 0.90 \\
		& & RTGEE  & 0.93 & 0.93 & 0.94& 0.0030 & 17.00 & 0.06 & 0.94 \\
		&  $R_{un}$ & SGEE & 0.96 & 0.92 & 0.94 & 0.0770 & 15.87 & 0.04 & 0.42 \\
		& & RSGEE & 0.93  & 0.95 & 0.96 & 0.0088 & 16.94 & 0.04 & 0.92 \\
		& & ERSGEE & 0.97 & 0.94  & 0.94 & 0.0046 & 16.91 & 0.00 & 0.91 \\
		& & RTGEE & 0.96 & 0.91 & 0.93 & 0.0032 & 17.00 & 0.07 & 0.93 \\
		\hline
	\end{tabular}%
	%	}
\end{table}

\begin{table}[H]
	\centering
	\caption{Correlated continuous data for large $n$ and diverging $p$ ($n=200$ and $p_n=[4n^{2/5}]-5$) with $\epsilon_{ij}$ following a t(3) distribution: Comparison of SGEE, RSGEE, ERSGEE, and the proposed method RTGEE with
		three different working correlation matrices (exchangeable, AR(1) and unstructured).}\label{table2-1}
	
	\small
	\resizebox{\textwidth}{!}{%
		\begin{tabular}{clrcccccccc}
			\hline			
			\multicolumn{3}{c}{ }&
			\multicolumn{1}{c}{$\beta_1$} &
			\multicolumn{1}{c}{$\beta_2$} &
			\multicolumn{1}{c}{$\beta_3$} &
			\multicolumn{1}{c}{ }&
			\multicolumn{2}{c}{No.of Zeros}\\
			\cline{4-6}\cline{8-9}
			Scenario & $R$ & Method & CI & CI& CI & MMSPE  & C & IC &CF\\
			\hline
			Case 1 & EXC & SGEE  & 0.93 & 0.95  & 0.93& 0.0301 & 21.58 & 0.06 & 0.46 \\
			& &  RSGEE& 0.96 & 0.95 & 0.94 & 0.0143 & 22.75 & 0.05 & 0.84 \\
			& & ERSGEE& 0.94 & 0.93 & 0.91 & 0.0158& 22.81 & 0.07 & 0.88 \\
			& & RTGEE & 0.97 & 0.95 & 0.89 & 0.0119 & 22.98 & 0.12 & 0.87 \\
			&   AR(1) & SGEE & 0.93 & 0.93 & 0.95 & 0.0274 & 21.48 & 0.05 & 0.45 \\
			& & RSGEE & 0.93 & 0.95 & 0.94 & 0.0143 & 22.82 & 0.06 & 0.85 \\
			& & ERSGEE & 0.96 & 0.95 & 0.95 & 0.0164 & 22.51 & 0.00 & 0.83 \\
			& & RTGEE & 0.93 & 0.92 & 0.96 & 0.0129 & 22.71 & 0.02 & 0.86 \\
			&  $R_{un}$  & SGEE & 0.94 & 0.92 & 1.00 & 0.0214 & 22.67 & 0.39 & 0.48 \\
			& & RSGEE & 0.95 & 0.95 & 0.94 & 0.0069& 22.98 & 0.06 & 0.92 \\
			& & ERSGEE & 0.93 & 0.92 & 0.94 & 0.0118 & 22.93 & 0.06 & 0.89 \\
			& & RTGEE & 0.95 & 0.94 & 0.97 & 0.0071 & 23.00 & 0.02 & 0.98 \\
			\hline
			Case 2 & EXC & SGEE & 0.97 & 0.99 & 0.98 & 0.3731 & 19.05 & 0.01 & 0.15 \\
			& & RSGEE & 0.96 & 0.97 & 0.98 & 0.0501 & 22.35 & 0.06 & 0.72 \\
			& & ERSGEE & 0.95 & 0.94 & 0.87 & 0.0218 & 22.94 & 0.12 & 0.84 \\
			& & RTGEE & 0.93 & 0.95 & 0.92 & 0.0193 & 22.93 & 0.08 & 0.87 \\
			&   AR(1) & SGEE & 0.97 & 0.93 & 0.95 & 0.3781 & 18.84 & 0.01 & 0.10 \\
			& & RSGEE & 0.92 & 0.96 & 0.94 & 0.0479 & 22.28 & 0.05 & 0.72 \\
			& & ERSGEE & 0.95 & 0.94 & 0.87 & 0.0213 & 22.91 & 0.12 & 0.84 \\
			& &RTGEE & 0.96 & 0.95 & 0.93 & 0.0201 & 22.77 & 0.05 & 0.84 \\
			&  $R_{un}$ & SGEE & 0.97 & 0.94 & 0.98 & 0.3429 & 18.82 & 0.03 & 0.13 \\
			& & RSGEE & 0.96 & 0.95 & 0.87 & 0.0352 & 22.93 & 0.12 & 0.81 \\
			& & ERSGEE & 0.96 & 0.92 & 0.88 & 0.0174 & 22.97 & 0.12 & 0.85 \\
			& & RTGEE  & 0.97 & 0.95 & 0.92 & 0.0146 & 22.93 & 0.07 & 0.86 \\
			\hline
			Case 3 & EXC& SGEE & 0.99 & 0.99 & 0.93 & 0.3062 & 18.86 & 0.02 & 0.10 \\
			& & RSGEE& 0.96 & 0.95 & 0.98 & 0.0341 & 22.69 & 0.08 & 0.79 \\
			& & ERSGEE & 0.97 & 0.95 & 0.85 & 0.0213 & 23.00 & 0.14 & 0.86 \\
			& & RTGEE & 0.97 & 0.97 & 0.92 & 0.0192 & 22.91 & 0.10 & 0.87 \\
			&   AR(1) & SGEE & 0.94 & 0.94 & 0.92 & 0.2950 & 18.82 & 0.01 & 0.08 \\
			& & RSGEE & 0.96 & 0.95 & 0.91 & 0.0337 & 22.72 & 0.08 & 0.81 \\
			& & ERSGEE & 0.98 & 0.92 & 0.86 & 0.0183 & 22.93 & 0.14 & 0.83 \\
			& & RTGEE & 0.97 & 0.96 & 0.94 & 0.0168 & 22.77 & 0.05 & 0.84 \\
			&  $R_{un}$ & SGEE & 0.94 & 0.96 & 0.93 & 0.3040 & 18.73 & 0.01 & 0.13 \\
			& & RSGEE & 0.95 & 0.94 & 0.87 & 0.0247 & 22.98 & 0.13 & 0.85 \\
			& & ERSGEE & 0.94 & 0.93 & 0.85 & 0.0141 & 23.00 & 0.14 & 0.86 \\
			& & RTGEE & 0.95  & 0.93 & 0.92 & 0.0114 & 22.97 & 0.08 & 0.89 \\
			\hline
		\end{tabular}%
	}
\end{table}

\begin{table}[H]
	\centering
	\caption{Correlated continuous data for $p>n$  ($n=100$ and $p=300$) with $\epsilon_{ij}$ following a t(3) distribution: Comparison of SGEE, RSGEE, ERSGEE, and the proposed method RTGEE with
		three different working correlation matrices (exchangeable, AR(1) and unstructured).}\label{table3-1}
	%	\noindent{\bf Table 3: Continuous data for $p>n$}\\
	%	{\small We set $n=100$, $m=10$, and $p=300$, and consider a linear model with $\epsilon_{i} \sim T_3(0,R(\alpha))$.}\label{tab1}
	\small
	\resizebox{\textwidth}{!}{%		
		\begin{tabular}{clccccccccc}
			\hline			
			\multicolumn{3}{c}{ }&
			\multicolumn{1}{c}{$\beta_1$} &
			\multicolumn{1}{c}{$\beta_2$} &
			\multicolumn{1}{c}{$\beta_3$} &
			\multicolumn{1}{c}{ }&
			\multicolumn{2}{c}{No.of Zeros}\\
			\cline{4-6}\cline{8-9}
			Scenario & $R$ & Method & CI & CI& CI & MMSPE  & C & IC &CF\\
			\hline
			Case 1 & EXC &SGEE & 0.93 & 0.95 & 0.96 & 0.0057 & 293.16 & 0.04 & 0.39 \\
			& & RSGEE & 0.94 & 0.96 & 0.97 & 0.0015& 296.83 & 0.03 & 0.82 \\
			& & ERSGEE & 0.95 & 0.94 & 0.95 & 0.0021 & 297.00 & 0.05 & 0.95 \\
			& & RTGEE & 0.95 & 0.94 & 0.96 & 0.0020 & 296.99 & 0.02 & 0.98 \\
			&   AR(1) &SGEE & 0.96 & 0.83 & 1.00 & 0.1159 & 295.53 & 0.54 & 0.30 \\
			& & RSGEE & 0.97 & 0.97 & 0.97 & 0.0023 & 296.80 & 0.03 & 0.83 \\
			& & ERSGEE & 0.95 & 0.94 & 0.97  & 0.0029 & 296.98 & 0.03 & 0.96 \\
			& & RTGEE & 0.95 & 0.96 & 0.97 & 0.0025 & 296.98 & 0.03 & 0.96 \\
			&  $R_{un}$  &SGEE & 0.94 & 0.96 & 0.95 & 0.0045 & 294.26 & 0.05 & 0.43 \\
			& & RSGEE & 0.94 & 0.97 & 0.97 & 0.0015 & 296.82 & 0.03 & 0.84 \\
			& & ERSGEE & 0.96 & 0.95 & 0.96 & 0.0026 & 296.98 & 0.04 & 0.95 \\
			& & RTGEE & 0.95 & 0.95 & 0.97 & 0.0023 & 296.98 & 0.03 & 0.96 \\
			\hline
			Case $2'$ &	EXC &SGEE & 0.95 & 0.96 & 0.89 & 0.3216 & 271.30 & 0.09 & 0.00 \\
			& & RSGEE & 0.96 & 0.95 & 0.97 & 0.0034 & 296.67 & 0.03 & 0.75 \\
			& & ERSGEE & 0.95 & 0.96 & 0.96 & 0.0027 & 296.90 & 0.04 & 0.88 \\
			& & RTGEE & 0.94 & 0.94 & 0.94 & 0.0027& 296.95 & 0.06 & 0.91 \\
			&   AR(1) &SGEE & 0.96 & 0.92 & 1.00 & 0.1445 & 289.53 & 0.49 & 0.02 \\
			& & RSGEE & 0.97 & 0.93 & 0.90 & 0.0050 & 296.85 & 0.10 & 0.78 \\
			& & ERSGEE & 0.97 & 0.90 & 0.87 & 0.0035 & 296.98 & 0.13 & 0.86 \\
			& &RTGEE & 0.97 & 0.91 & 0.88 & 0.0034 & 296.98 & 0.12 & 0.87 \\
			&  $R_{un}$ & SGEE & 0.97 & 0.95 & 0.90 & 0.3642 & 271.55 & 0.07 & 0.01 \\
			& & RSGEE & 0.96 & 0.95 & 0.96 & 0.0039 & 296.66 & 0.04 & 0.73 \\
			& & ERSGEE & 0.97 & 0.95 & 0.95 & 0.0031 & 296.90 & 0.05 & 0.87 \\
			& & RTGEE & 0.94 & 0.95 & 0.95& 0.0028 & 296.89 & 0.05 & 0.86 \\
			\hline
			Case $3'$ &	EXC & SGEE & 0.96 & 0.96 & 0.94 & 0.6168 & 242.35 & 0.00 & 0.01 \\
			& &  RSGEE & 0.94 & 0.95 & 0.94 & 0.0042 & 296.75 & 0.06 & 0.79 \\
			& & ERSGEE & 0.93 & 0.92 & 0.93 & 0.0026 & 296.87 & 0.07 & 0.86 \\
			& & RTGEE & 0.94 & 0.93 & 0.95 & 0.0026 & 296.87 & 0.05 & 0.88 \\
			&   AR(1) & SGEE & 0.94 & 0.95 & 0.93 & 0.7165 & 242.15 & 0.00 & 0.00 \\
			& & RSGEE & 0.94 & 0.95 & 0.93 & 0.0052 & 296.78 & 0.07 & 0.79 \\
			& & ERSGEE & 0.95 & 0.93 & 0.93 & 0.0028 & 296.87 & 0.07 & 0.86 \\
			& & RTGEE & 0.96 & 0.93& 0.93  & 0.0030 & 296.87 & 0.07 & 0.86 \\
			&  $R_{un}$ & SGEE & 0.93 & 0.95& 0.96 & 0.6998 & 242.27 & 0.00 & 0.00 \\
			& & RSGEE & 0.94 & 0.94 & 0.93 & 0.0039 & 296.79 & 0.06 & 0.80 \\
			& & ERSGEE & 0.94 & 0.93 & 0.93 & 0.0029 & 296.88 & 0.07 & 0.87 \\
			& & RTGEE & 0.94 & 0.94 & 0.95 & 0.0027 & 296.87 & 0.05 & 0.88 \\
			\hline
		\end{tabular} %
	}
\end{table}

\begin{table}[H]
	\centering
	\caption{Correlated continuous data  with $\epsilon_{ij}$ following a normal distribution for $n>p$ ($p=20$ and $n=100$):
		Comparison of SGEE, RSGEE, ERSGEE, and the proposed method RTGEE with
		three different working correlation matrices (exchangeable, AR(1) and unstructured).}\label{table4-1}
	%	\noindent{\bf Table 1: Continuous data for $n>p$}\\
	%	{\small We consider linear model with $\epsilon_{i}$
	%		following a multivariate distribution.}\label{tab1}  %heavy tail t(3) distribution
	\small
	\resizebox{\textwidth}{!}{%
		\begin{tabular}{clrcccccccc}
			\hline			
			\multicolumn{3}{c}{ }&
			\multicolumn{1}{c}{$\beta_1$} &
			\multicolumn{1}{c}{$\beta_2$} &
			\multicolumn{1}{c}{$\beta_3$} &
			\multicolumn{1}{c}{ }&
			\multicolumn{2}{c}{No.of Zeros}\\
			\cline{4-6}\cline{8-9}
			Scenario & $R$ & Method & CI & CI& CI & MMSPE  & C & IC &CF\\
			\hline
			Case 1 & EXC & SGEE & 0.94 & 0.98 & 0.98 & 0.0008 & 16.99 & 0.02 & 0.97 \\
			& &  RSGEE & 0.96 & 0.98 & 0.98 & 0.0008 & 16.97 & 0.02 & 0.95 \\
			& & ERSGEE & 0.95 & 0.97 & 0.96 & 0.0008 & 16.99 & 0.00 & 0.99 \\
			& & RTGEE  & 0.96 & 0.98 & 0.98 & 0.0009 & 17.00 & 0.02 & 0.98 \\
			&   AR(1) & SGEE & 0.96 & 0.97 & 0.99 & 0.0009 & 16.90 & 0.01 & 0.91 \\
			& & RSGEE & 0.95 & 0.96 & 0.96 & 0.0011 & 16.97 & 0.04 & 0.93 \\
			& & ERSGEE & 0.95 & 0.93 & 0.95 & 0.0013 & 17.00 & 0.00 & 1.00 \\
			& & RTGEE & 0.94 & 0.98 & 0.98 & 0.0011 & 17.00 & 0.01 & 0.99 \\
			&  $R_{un}$  & SGEE & 0.93 & 0.99 & 0.99 & 0.0009 & 16.97 & 0.01 & 0.96 \\
			& & RSGEE  & 0.96 & 0.98 & 0.98 & 0.0010 & 16.96 & 0.02 & 0.94 \\
			& & ERSGEE  & 0.95 & 0.98 & 0.95 & 0.0010 & 17.00 & 0.00 & 1.00 \\
			& & RTGEE & 0.95 & 0.98 & 0.98 & 0.0009 & 17.00 & 0.02 & 0.98 \\
			\hline
			Case 2 & EXC & SGEE & 0.95 & 0.94 & 0.91 & 0.0528 & 16.75 & 0.05 & 0.75 \\
			& & RSGEE & 0.96 & 0.94 & 0.98 & 0.0046 & 16.94 & 0.01 & 0.93 \\
			& & ERSGEE & 0.96 & 0.94 & 0.96 & 0.0017 & 16.99 & 0.00 & 0.99 \\
			& & RTGEE & 0.96 & 0.95 & 0.98 & 0.0014 & 17.00 & 0.02 & 0.98 \\
			&   AR(1) & SGEE & 0.97 & 0.94 & 1.00 & 0.0779 & 16.94 & 0.27 & 0.67 \\
			& & RSGEE & 0.96 & 0.95 & 0.97 & 0.0057 & 16.95 & 0.02 & 0.93 \\
			& & ERSGEE & 0.92 & 0.96 & 0.93 & 0.0022 & 16.99 & 0.00 & 0.99 \\
			& &RTGEE & 0.94 & 0.96 & 0.96 & 0.0021 & 17.00 & 0.02 & 0.98 \\
			&  $R_{un}$ & SGEE & 0.96 & 0.96 & 0.92 & 0.0633 & 16.81 & 0.04 & 0.79 \\
			& & RSGEE & 0.97 & 0.97 & 0.96 & 0.0051 & 16.96 & 0.01 & 0.95 \\
			& & ERSGEE & 0.93 & 0.95 & 0.96 & 0.0018 & 16.99 & 0.00 & 0.99 \\
			& & RTGEE & 0.94 & 0.97& 0.98 & 0.0016 & 16.99 & 0.02 & 0.97 \\
			\hline
			Case 3 & EXC & SGEE & 0.95& 0.97 & 0.94 & 0.0537 & 16.66 & 0.04 & 0.75 \\
			& &  RSGEE & 0.95& 0.97& 0.96 & 0.0047 & 16.93 & 0.00 & 0.94 \\
			& & ERSGEE & 0.94 & 0.97 & 0.95 & 0.0024 & 16.91 & 0.00 & 0.93 \\
			& & RTGEE & 0.96 & 0.95 & 0.95 & 0.0017 & 16.99 & 0.05 & 0.94 \\
			&   AR(1) & SGEE & 0.92 & 0.98 & 0.94 & 0.0583 & 16.59 & 0.04 & 0.69 \\
			& & RSGEE & 0.95 & 0.95 & 0.94 & 0.0048 & 16.86 & 0.00 & 0.88 \\
			& & ERSGEE & 0.94 & 0.93 & 0.95 & 0.0024 & 16.90 & 0.00 & 0.90 \\
			& & RTGEE & 0.94 & 0.95 & 0.95 & 0.0022 & 16.96 & 0.05 & 0.92 \\
			&  $R_{un}$ & SGEE & 0.95 & 0.96 & 0.93 & 0.0577 & 16.56 & 0.04 & 0.68 \\
			& & RSGEE & 0.96 & 0.95 & 0.96 & 0.0055 & 16.88 & 0.00 & 0.89 \\
			& & ERSGEE & 0.95 & 0.94 & 0.95 & 0.0026 & 16.90 & 0.00 & 0.90 \\
			& & RTGEE & 0.95 & 0.96 & 0.96 & 0.0022 & 16.94 & 0.04 & 0.90 \\
			\hline
		\end{tabular}%
	}
\end{table}

\begin{table}[H]
	\centering
	\caption{Correlated continuous data for large $n$ and diverging $p$ ($n=200$ and $p_n=[4n^{2/5}]-5$) with $\epsilon_{ij}$ following a normal distribution:
		Comparison of SGEE, RSGEE, ERSGEE, and the proposed method RTGEE with
		three different working correlation matrices (exchangeable, AR(1) and unstructured).}\label{table5-1}
	%	\noindent{\bf Table 2: Continuous data with diverging $p$}\\
	%	{\small We set $n=200$, $m \in \{2,3,4,5\}$, $p_n=[4n^{2/5}]-5$, and $s_n=[p_n/5]$. We consider linear model with $\epsilon_{i}$
	%		following a multivariate t distribution with three degrees of freedom.}\label{tab1}
	\small
	\resizebox{\textwidth}{!}{%
		\begin{tabular}{clrcccccccc}
			\hline			
			\multicolumn{3}{c}{ }&
			\multicolumn{1}{c}{$\beta_1$} &
			\multicolumn{1}{c}{$\beta_2$} &
			\multicolumn{1}{c}{$\beta_3$} &
			\multicolumn{1}{c}{ }&
			\multicolumn{2}{c}{No.of Zeros}\\
			\cline{4-6}\cline{8-9}
			Scenario & $R$ & Method & CI & CI& CI & MMSPE  & C & IC &CF\\
			\hline
			Case 1 & EXC & SGEE & 0.98 & 0.99  & 0.95 & 0.0068 & 22.48 & 0.03 & 0.69 \\
			& &  RSGEE & 0.93 & 0.95 & 0.99 & 0.0060 & 22.83 & 0.01 & 0.90 \\
			& & ERSGEE & 0.96 & 0.96 & 0.95 & 0.0111 & 22.93 & 0.02 & 0.95 \\
			& & RTGEE & 0.93 & 0.98 & 0.94 & 0.0061 & 22.97 & 0.02 & 0.95 \\
			&   AR(1) &SGEE & 0.93 & 0.92 & 0.98 & 0.0069 & 22.47 & 0.02 & 0.69 \\
			& & RSGEE & 0.92 & 0.93 & 0.98 & 0.0065 & 22.83 & 0.02 & 0.89 \\
			& & ERSGEE & 0.95 & 0.96 & 0.93 & 0.0117 & 22.94 & 0.01 & 0.94 \\
			& & RTGEE & 0.92 & 0.94 & 0.98 & 0.0059 & 22.94 & 0.02 & 0.95 \\
			&  $R_{un}$  &SGEE & 0.93 & 0.96 & 0.99 & 0.0039 & 22.64 & 0.01 & 0.76 \\
			& & RSGEE & 0.94 & 0.94 & 0.99 & 0.0039 & 22.98 & 0.01 & 0.97 \\
			& & ERSGEE & 0.96 & 0.96 & 0.95 & 0.0079 & 22.91 & 0.05 & 0.94 \\
			& & RTGEE & 0.94 & 0.96 & 0.98 & 0.0038 & 22.98 & 0.01 & 0.97 \\
			\hline
			Case 2 & EXC & SGEE & 0.95 & 0.99 & 0.99 & 0.2582 & 21.37 & 0.04 & 0.36 \\
			& & RSGEE & 0.95 & 0.97 & 0.97& 0.0232 & 22.69 & 0.03 & 0.81 \\
			& & ERGEE & 0.93 & 0.99 & 0.94 & 0.0101 & 22.92 & 0.05 & 0.91 \\
			& & RTGEE & 0.93  & 0.93 & 0.94 & 0.0088 & 22.93 & 0.04 & 0.91 \\
			&   AR(1) & SGEE & 0.96 & 0.94 & 0.95 & 0.2511 & 21.17 & 0.01 & 0.35 \\
			& & RSGEE & 0.93 & 0.96& 0.96 & 0.0248 & 22.62 & 0.04 & 0.77 \\
			& & ERSGEE & 0.95 & 0.95 & 0.95 & 0.0096 & 22.95 & 0.05 & 0.91 \\
			& &RTGEE & 0.94 & 0.94 & 0.96 & 0.0122 & 22.91 & 0.04 & 0.92 \\
			&  $R_{un}$ & SGEE & 0.95 & 0.93 & 0.97 & 0.1980 & 21.32 & 0.02 & 0.41 \\
			& & RSGEE & 0.94 & 0.95 & 0.96 & 0.0154 & 22.99 & 0.04 & 0.95 \\
			& & ERSGEE & 0.98 & 0.95 & 0.94 & 0.0088 & 23.00 & 0.06 & 0.94 \\
			& & RTGEE & 0.96 & 0.93  & 0.94 & 0.0061 & 23.00 & 0.06 & 0.94 \\
			\hline
			Case 3 & EXC & SGEE & 0.99 & 0.99 & 0.99 & 0.2307 & 21.09 & 0.03 & 0.23 \\
			& & RSGEE  & 0.96 & 0.99 & 0.93 & 0.0205 & 22.63 & 0.03 & 0.84 \\
			& & ERSGEE & 0.97 & 0.94 & 0.92 & 0.0094 & 23.00 & 0.07 & 0.93 \\
			& & RTGEE & 0.95 & 0.96 & 0.94 & 0.0084 & 22.99 & 0.06 & 0.93 \\
			&   AR(1) & SGEE & 0.95 & 0.96 & 0.94 & 0.2386 & 21.09 & 0.04 & 0.25 \\
			& & RSGEE & 0.94 & 0.95 & 0.93 & 0.0200 & 22.71 & 0.03 & 0.82 \\
			& & ERSGEE & 0.94 & 0.92 & 0.94 & 0.0087 & 23.00 & 0.06 & 0.94 \\
			& & RTGEE & 0.95 & 0.92 & 0.94 & 0.0083 & 23.00 & 0.06 & 0.94 \\
			&  $R_{un}$ & SGEE & 0.95 & 0.96 & 0.95 & 0.2146 & 21.07 & 0.04 & 0.25 \\
			& & RSGEE & 0.94 & 0.97 & 0.95 & 0.0146 & 22.97 & 0.03 & 0.94 \\
			& & ERSGEE & 0.95 & 0.93 & 0.93 & 0.0067 & 23.00 & 0.07 & 0.93 \\
			& & RTGEE & 0.95 & 0.96 & 0.96 & 0.0061 & 23.00 & 0.04 & 0.96 \\
			\hline
		\end{tabular}%
	}
\end{table}

\begin{table}[H]
	\centering
	\caption{Correlated continuous data for $p>n$  ($n=100$ and $p=300$) with $\epsilon_{ij}$ following a normal distribution:
		Comparison of SGEE, RSGEE, ERSGEE, and the proposed method RTGEE with
		three different working correlation matrices (exchangeable, AR(1) and unstructured).}\label{table6-1}
	\small
	\resizebox{\textwidth}{!}{%		
		\begin{tabular}{clrcccccccc}
			\hline			
			\multicolumn{3}{c}{ }&
			\multicolumn{1}{c}{$\beta_1$} &
			\multicolumn{1}{c}{$\beta_2$} &
			\multicolumn{1}{c}{$\beta_3$} &
			\multicolumn{1}{c}{ }&
			\multicolumn{2}{c}{No.of Zeros}\\
			\cline{4-6}\cline{8-9}
			Scenario & $R$ & Method & CI & CI& CI & MMSPE  & C & IC &CF\\
			\hline
			Case 1 & EXC &SGEE & 0.96 & 0.95 & 0.95 & 0.0012 & 296.51 & 0.05 & 0.75 \\
			& & RSGEE & 0.94 & 0.97 & 0.97 & 0.0012 & 296.90 & 0.03 & 0.89 \\
			& & ERSGEE & 0.96 & 0.98 & 0.98 & 0.0023 & 297.00 & 0.02 & 0.98 \\
			& & RTGEE & 0.95 & 0.98 & 0.97 & 0.0011 & 297.00 & 0.02 & 0.98 \\
			&   AR(1) &SGEE & 0.96 & 0.98 & 0.98 & 0.0017 & 296.36 & 0.02 & 0.72 \\
			& & RSGEE & 0.97 & 0.95 & 0.95 & 0.0017 & 296.86 & 0.05 & 0.87 \\
			& & ERSGEE & 0.93 & 0.97 & 0.98 & 0.0032 & 297.00 & 0.02 & 0.98 \\
			& & RTGEE & 0.95 & 0.99 & 0.98 & 0.0016 & 297.00 & 0.01 & 0.99 \\
			&  $R_{un}$  &SGEE & 0.96 & 0.97 & 0.97 & 0.0011 & 296.55 & 0.03 & 0.76 \\
			& & RSGEE & 0.95 & 0.96 & 0.96 & 0.0011 & 296.89 & 0.04 & 0.87 \\
			& & ERSGEE & 0.96 & 0.93 & 0.93 & 0.0023 & 297.00 & 0.07 & 0.93 \\
			& & RTGEE & 0.96 & 0.94& 0.94 & 0.0010 & 297.00 & 0.06 & 0.94 \\
			\hline
			Case $2''$ & EXC &SGEE & 0.95 & 0.94 & 0.98 & 0.0566 & 283.38 & 0.02 & 0.26 \\
			& & RSGEE & 0.94 & 0.96 & 0.99 & 0.0020 & 296.81 & 0.01 & 0.87 \\
			& & ERSGEE & 0.94 & 0.96 & 0.97 & 0.0018 & 296.99 & 0.03 & 0.96 \\
			& & RTGEE & 0.95 & 0.96 & 0.98 & 0.0014  & 296.97 & 0.02 & 0.96 \\
			&   AR(1) &SGEE & 0.96 & 0.95 & 0.98 & 0.0374 & 277.70 & 0.02 & 0.25 \\
			& & RSGEE & 0.96 & 0.95 & 0.97 & 0.0023 & 296.77 & 0.03 & 0.82 \\
			& & ERSGEE & 0.96 & 0.95 & 0.97 & 0.0024 & 296.99 & 0.03 & 0.96 \\
			& &RTGEE & 0.98 & 0.97 & 0.98 & 0.0019 & 296.97 & 0.02 & 0.95 \\
			&  $R_{un}$ & SGEE & 0.94 & 0.95 & 0.97 & 0.0546 & 276.36 & 0.03 & 0.24 \\
			& & RSGEE & 0.93 & 0.96 & 0.98 & 0.0019 & 296.84 & 0.02 & 0.87 \\
			& & ERSGEE & 0.94 & 0.96 & 0.97 & 0.0022 & 296.99 & 0.03 & 0.96 \\
			& & RTGEE & 0.94 & 0.97 & 0.97 & 0.0018 & 296.98 & 0.03 & 0.95 \\
			\hline
			Case $3''$ & EXC & SGEE & 0.96 & 0.92 & 0.91 & 0.0549 & 282.37 & 0.09 & 0.18 \\
			& &  RSGEE & 0.97 & 0.94 & 0.96 & 0.0024 & 296.75 & 0.04 & 0.78 \\
			& & ERSGEE & 0.93 & 0.95 & 0.96 & 0.0020 & 296.97 & 0.04 & 0.94 \\
			& & RTGEE & 0.96 & 0.96 & 0.96 & 0.0016 & 296.97 & 0.04 & 0.94 \\
			&   AR(1) & SGEE & 0.95 & 0.91 & 0.90 & 0.0407 & 286.46 & 0.09 & 0.23 \\
			& & RSGEE & 0.96 & 0.93 & 0.96 & 0.0033 & 296.82 & 0.04 & 0.84 \\
			& & ERSGEE & 0.97 & 0.95 & 0.96 & 0.0027 & 296.97 & 0.04 & 0.94 \\
			& & RTGEE  & 0.97 & 0.95 & 0.96 & 0.0022 & 296.97 & 0.04 & 0.94 \\
			&  $R_{un}$ & SGEE & 0.95 & 0.93 & 0.90 & 0.0653 & 281.18 & 0.10 & 0.19 \\
			& & RSGEE & 0.95 & 0.94 & 0.96 & 0.0028 & 296.86 & 0.04 & 0.84 \\
			& & ERSGEE & 0.93 & 0.95 & 0.96 & 0.0025 & 296.97 & 0.04 & 0.94 \\
			& & RTGEE & 0.94 & 0.96 & 0.96 & 0.0020 & 296.97 & 0.04 & 0.94 \\
			\hline
		\end{tabular} %
	}
\end{table}

\begin{table}[H]
	\centering
	\caption{The parameter estimates of selected TFs, the mean squared error
		for cross validation procedures under three correlation structures, and the running time (s means seconds) for four procedures in the yeast cell-cycle process.}\label{table7}
	%	\noindent{\bf Table 6: Yeast Cell-Cycle Gene Expression Data Analysis}\\
	%	{\small The parameter estimates of selected TFs under three correlation structures for four procedures. Number of TFs selected for G1 stage in the yeast cell-cycle process, the mean squared error
	%		for cross validation procedures, and lists of selected TFs under three correlation structures for four procedures.}\label{tab1}
	\small
	\resizebox{\textwidth}{!}{%
		\begin{tabular}{rrrrrrrrrrrrrr}
			\hline			
			\multicolumn{1}{c}{ }&
			\multicolumn{3}{c}{SGEE} &
			\multicolumn{3}{c}{RSGEE} &
			\multicolumn{3}{c}{ERSGEE} &
			\multicolumn{3}{c}{RTGEE} \\
			\cline{2-4}\cline{5-7}\cline{8-10}\cline{11-13}
			Covariates & EXC & AR(1) & $R_{un}$ & EXC & AR(1) & $R_{un}$ & EXC & AR(1) & $R_{un}$ & EXC & AR(1) & $R_{un}$ \\
			\hline
			intercept &  0.098 &  0.105 &  0.068 &  0.113 &  0.121 &0.098 & 0.119 & 0.134 & 0.099 & 0.126 & 0.137 & 0.087\\
			time &  0.010 &  0.008 &  0.010 &  0.007 &  0.006 &0.008 & 0.006 & 0.004 &0.008 & 0.006 & 0.004 & 0.009\\
			ABF1 & -0.048 & -0.047 & -0.045 &     0   &  0      &0  & 0 & 0 &0 & 0 & 0 & 0\\
			ACE2 &  0.041 &  0.041 &  0.045 &     0   &  0      &0 & 0 & 0 &0 & 0 & 0 & 0 \\
			ASH1 & -0.104 & -0.094 & -0.073 & -0.113 & -0.107 &-0.101 & -0.124 & -0.123 &-0.099 & -0.125 & -0.117& -0.092\\
			CIN5 &  0.044 &  0.048 &  0.059 &     0   &  0      &0  & 0 & 0 &0 & 0 & 0 & 0\\
			CUP9 & -0.061 & -0.050 & -0.028 & -0.058 & -0.047 &-0.042 & -0.064 & -0.055 &-0.045 & -0.057 & -0.048& -0.035 \\
			FKH2 & -0.111 & -0.106 & -0.097 & -0.110 & -0.102 &-0.094 & -0.117 & -0.108 &-0.092 & -0.119 & -0.106 & -0.092\\
			GAL4 & -0.035 & -0.020 & -0.009 &    0    &  0      &0 & 0 & 0 &0 & 0 & 0 & 0\\
			GAT3 &  0.493 &  0.459 &  0.436 &  0.434 &  0.422 &0.385 & 0.441 & 0.411 &0.389 & 0.443 & 0.427 & 0.399\\
			GCR1 & -0.071 & -0.068 & -0.066 & -0.056 & -0.056 &-0.054  & -0.051 & -0.053 &-0.055 & -0.053 & -0.056 & -0.056 \\
			GCR2 & -0.098 & -0.086 & -0.071 &  0.001 &  0.011 &0.026  & 0.019 & 0.029 &0.021 & 0.002 & 0.009 & 0.011\\
			GLN3 &  0.033 &  0.040 &  0.049 & -0.008 & -0.002 &0.005  & -0.005 & -0.000 &0.004 & -0.011 & -0.004 & 0 \\
			GRF10.Pho2   & -0.035 & -0.035 & -0.037 & -0.018 & -0.016 &-0.008  & -0.007 & -0.001 & -0.009 & -0.009 & -0.009 & -0.017\\
			HAP2 & -0.179 & -0.166 & -0.079 & -0.525 & -0.466 &-0.395  & -0.454 & -0.404 &-0.295 & -0.641 & -0.548& -0.478 \\
			HAP3 & -0.084 & -0.082 & -0.080 & -0.032 & -0.031 &-0.029 & -0.031 & -0.030 &-0.029  & -0.031 & -0.030 & -0.029 \\
			IME4 &  0.142 &  0.134 &  0.057 &  0.498 &  0.444 &0.375 & 0.422 & 0.377 &0.273 & 0.614 & 0.527 & 0.462\\
			IXR1 & -0.059 & -0.059 & -0.060 &  0      &  0      & 0& 0  & 0 &0 & 0 & 0 & 0\\
			MAC1 & -0.022 & -0.018 & -0.013 & -0.012 & -0.010 &-0.009 & -0.016& -0.015 &-0.011 & -0.011 & -0.009 & -0.006 \\
			MBP1 &  0.106 &  0.099 &  0.083 &  0.133 &  0.125 &0.117  & 0.134 & 0.124 &0.119 & 0.134 & 0.126 & 0.118 \\
			MET31& -0.081 & -0.080 & -0.072 & -0.072 & -0.069 &-0.067 & -0.078 & -0.079 &-0.063 & -0.080 & -0.076 & -0.064 \\
			MET4 & -0.058 & -0.058 & -0.062 & -0.049 & -0.045 &-0.030 & -0.032 & -0.020 &-0.026& -0.030 & -0.019 & -0.043 \\
			MTH1 & -0.024 & -0.023 & -0.024 &   0     &  0 &0 & 0 & 0 &0 & 0 & 0 & 0\\
			NDD1 & -0.100 & -0.101 & -0.107 & -0.089 & -0.095 &-0.104 & -0.073 & -0.084 &-0.107 & -0.072 & -0.090 & -0.110\\
			NRG1 & 0.073  & 0.063  &  0.040 &  0.060 &  0.047 &0.041 & 0.069 & 0.056 &0.044 & 0.059& 0.045 & 0.032 \\
			PDR1 & 0.150  & 0.119  &  0.091 &  0.109 &  0.101 &0.091 & 0.124 & 0.123 &0.092 & 0.114 & 0.105 & 0.085 \\
			ROX1 & 0.094  & 0.089  &  0.079 &  0.077 &  0.072 &0.067 & 0.075 & 0.072 &0.070 & 0.076 & 0.075 & 0.070 \\
			RTG3 & 0.064  & 0.065  &  0.067 &     0   &  0 &0 & 0  & 0 &0 & 0 & 0 & 0\\
			SRD1 & 0.056  & 0.047  &  0.039 & -0.043 & -0.050 &-0.063 & -0.060 & -0.067 &-0.059 & -0.043 & -0.047 & -0.047 \\
			STB1 & 0.103  & 0.103  &  0.102 &  0.095 &  0.096 &0.100 & 0.090 & 0.095 &0.096 & 0.090 & 0.091 & 0.096 \\
			STP1 & 0.080  & 0.076  &  0.069 &  0.100 &  0.099 &0.093 & 0.100 & 0.098 &0.091 & 0.100 & 0.096 & 0.094 \\
			SWI4 & 0.051  & 0.049  &  0.045 &  0.066 &  0.069 &0.073 & 0.069 & 0.077 &0.070 & 0.070 & 0.074 & 0.068 \\
			SWI6 & 0.064  & 0.062  &  0.057 &  0.040 &  0.035 &0.030 & 0.039 & 0.033 & 0.029 & 0.039 & 0.031 & 0.029 \\
			YAP5 &-0.477  &-0.439  & -0.416 & -0.406 & -0.394 &-0.358  & -0.416 & -0.388 & -0.364 & -0.415 & -0.401 & -0.374 \\
			YAP6 &-0.053  &-0.055  & -0.061 &    0    &  0      &0 & 0 & 0 & 0 & 0 & 0 & 0\\
			ZAP1 &-0.039  &      0  &     0   &    0    &  0      &0 & 0& 0 & 0 & 0 & 0 & 0\\
			\hline
			$\mathrm{MSE}_{\mathrm{CV}}$ & 1.780& 1.802& 1.847 & 2.232 & 2.202 & 2.186 & 2.070 & 2.042 & 1.969 & 1.862 &1.837 & 1.858     \\
			\hline
			Time (s) & 2.417  &2.270   &3.604   &11.011   &11.434   &20.251  &48.713   &49.774   &85.775   &40.735   & 40.044   &73.160  \\
			\hline
		\end{tabular} %
	}
\end{table}


\label{lastpage}

\end{document}